\documentclass[prd,12pt]{article}

\usepackage{tikz}
\usepackage{amsmath,amssymb,graphicx,multirow,xspace,slashed,array,booktabs}
\usepackage[colorlinks=true,urlcolor=blue,anchorcolor=blue,citecolor=blue,filecolor=blue,linkcolor=blue,menucolor=blue,pagecolor=blue]{hyperref}
\usepackage[compress,numbers]{natbib}
\usepackage{subfigure, placeins}
\usepackage{booktabs}
\usepackage{blindtext, rotating}
\usepackage{afterpage}
\usepackage{enumitem}
\usepackage{ marvosym }
\usepackage{authblk} 
\usepackage{verbatim}
\usepackage{soul} 
\usepackage[normalem]{ulem}
\usepackage{pifont}
\newcommand{\cmark}{\text{\textcolor{darkgreen}{\ding{51}}}}%
\newcommand{\xmark}{\text{\textcolor{red}{\ding{55}}}}%

\usepackage[utf8]{inputenc}

\usepackage{floatrow}
\newfloatcommand{capbtabbox}{table}[][\FBwidth]

\usepackage[font=footnotesize,labelfont=bf]{caption}

\usepackage{lineno}

\allowdisplaybreaks

\long\def\symbolfootnote[#1]#2{\begingroup%
\def\thefootnote{\fnsymbol{footnote}}\footnote[#1]{#2}\endgroup}

\newcommand{\newc}{\newcommand}
\newc{\gsim}{\lower.7ex\hbox{$\;\stackrel{\textstyle>}{\sim}\;$}}
\newc{\lsim}{\lower.7ex\hbox{$\;\stackrel{\textstyle<}{\sim}\;$}}
\newc{\gev}{\,{\rm GeV}}
\newc{\mev}{\,{\rm MeV}}
\newc{\ev}{\,{\rm eV}}
\newc{\kev}{\,{\rm keV}}
\newc{\tev}{\,{\rm TeV}}
\newc{\MHT}{$H_T^{\text{miss}}$}
\newc{\MET}{$\slashed{E}_T$}
\newc{\MTT}{$M_{T2}$}

\newc{\mz}{M_Z}
\newc{\mpl}{M_*}
\newc{\mw}{m_{\rm weak}}
\newc{\nr}[1]{N^c_R{}_{#1}}

\definecolor{darkgreen}{rgb}{0,0.5,0}
\definecolor{goodorange}{rgb}{0.9,0.4,0}

\def\beq{\begin{equation}}
\def\eeq{\end{equation}}
\newcommand{\bea}{\begin{eqnarray}\begin{aligned}}
\newcommand{\eea}{\end{aligned}\end{eqnarray}}
\def\bitem{\begin{itemize}}
\def\eitem{\end{itemize}}

\numberwithin{equation}{section}

\newcommand\fverb{\setbox\fverbbox=\hbox\bgroup\verb}

\newbox\fverbbox

\begin{document}

\baselineskip 0.6cm

\begin{titlepage}

\thispagestyle{empty}

\begin{center}

\vskip 1cm

{\Huge \bf Asymmetry Observables and the}\vskip0.5cm
{\Huge\bf Origin of $R_{D^{(*)}}$ Anomalies}

\vskip 1cm

\vskip 1.0cm
{\large Pouya Asadi, Matthew~R.~Buckley, and David Shih }
\vskip 1.0cm
{\it  NHETC, Dept.~of Physics and Astronomy\\ Rutgers, The State University of NJ \\ Piscataway, NJ 08854 USA} \\
\vskip 1.0cm

\end{center}

\vskip 1cm

\begin{abstract}

The $R_{D^{(*)}}$ anomalies are among the longest-standing and most statistically significant hints of physics beyond the Standard Model. Many models have been proposed to explain these anomalies, including the interesting possibility that right-handed neutrinos could be involved in the $B$ decays. In this paper, we investigate future measurements at Belle II that can be used to tell apart the various new physics scenarios.  Focusing on  a number of $\tau$ asymmetry observables (forward-backward asymmetry and polarization asymmetries) which can be reconstructed at Belle II, we calculate the contribution of the most general dimension 6 effective Hamiltonian (including right-handed neutrinos) to all of these asymmetries. We show that Belle~II can use these asymmetries to distinguish between new-physics scenarios that use right- and left-handed neutrinos, and in most cases can likely distinguish the specific model itself.

\end{abstract}

\flushbottom

\end{titlepage}

\setcounter{page}{1}

\tableofcontents

\vfill\eject

\section{Introduction}
\label{sec:intro}

Among the most tantalizing hints of new physics (NP) currently are a number of flavor anomalies \cite{Aubert:2007dsa, Bozek:2010xy,Lees:2012xj, Lees:2013uzd, Aaij:2015yra,Huschle:2015rga, Abdesselam:2016xqt,Aaij:2017vbb,Aaij:2017tyk}. Of these, one of the largest and longest-standing statistical discrepancies with the Standard Model (SM) is observed in the decays $B \rightarrow D^{(*)} \tau \nu$. This can be seen in the ratios $R_{D}$ and $R_{D^*}$, defined as
\begin{equation}
R_D = \frac{\Gamma (\bar{B} \rightarrow D \tau \nu)}{\Gamma (\bar{B} \rightarrow D \ell \nu)}, \hspace{0.4in}  R_{D^{*}} = \frac{\Gamma (\bar{B} \rightarrow D^{*} \tau \nu)}{\Gamma (\bar{B} \rightarrow D^{*} \ell \nu)},
\label{eq:rddef}
\end{equation}
where $\ell$ stands for either electrons or muons. The current global average \cite{HFLAV16} of the observed values are 
\begin{equation}
R_D = 0.407 \pm 0.046 , \hspace{0.5 in}  R_{D^{*}} = 0.304 \pm 0.015,
\label{eq:rdobs}
\end{equation}
while the Standard Model predictions are \cite{Lees:2012xj,Lees:2013uzd,HFLAV16,Kamenik:2008tj,Fajfer:2012vx,Bailey:2012jg,Lattice:2015rga,Na:2015kha,Aoki:2016frl,Amhis:2016xyh,Jaiswal:2017rve} 
\begin{equation}
R_D =0.299 \pm 0.003 , \hspace{0.5 in}  R_{D^{*}} = 0.258 \pm 0.005.
\label{eq:rdsm}
\end{equation}  
A combined analysis \cite{Amhis:2016xyh} shows a $\sim 3.8\sigma$ discrepancy \cite{HFLAV16} between the experimental results~\eqref{eq:rdobs} and the SM predictions~\eqref{eq:rdsm}.

\begin{figure}
\includegraphics[scale=1]{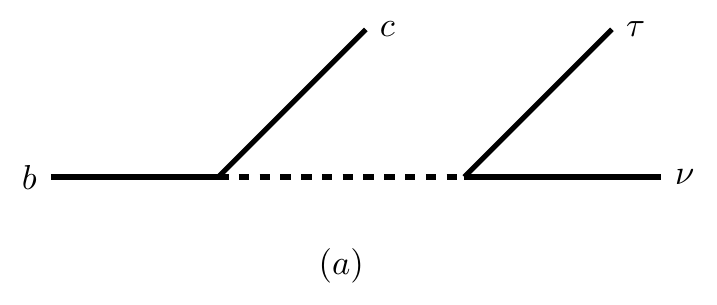}
\includegraphics[scale=1]{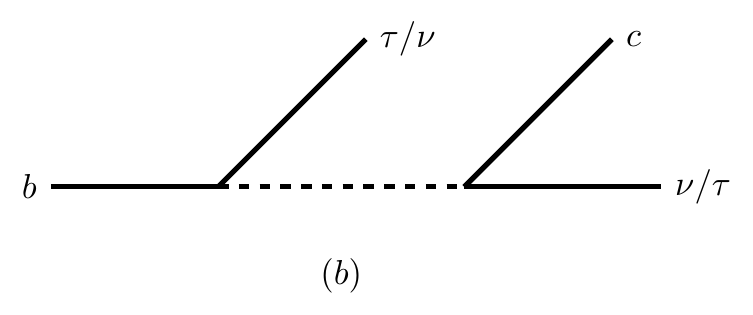}
\caption{The diagrams that modify the $b \rightarrow c \tau \nu$ rate, and subsequently $R_D$ and $R_{D^*}$, with a new BSM mediator. The mediator can be one of the three types indicated in the text: (a) uncolored mediators: charged scalar or $W'$; or (b) colored mediators: leptoquarks.}
\label{fig:diagram}
\end{figure}

Theoretical models proposed to explain these anomalies rely on new heavy mediators which enhance the $\bar{B} \rightarrow D^{(*)} \tau \nu$ decay rate. These mediators can be classified  by their spin (scalar or vector) and by whether they carry $SU(3)$ color. The possibilities essentially boil down to three categories: a colorless charged scalar (possibly part of an extended Higgs sector), a heavy charged vector boson ($W'$), or various types of scalar and vector leptoquarks (LQs). The generic tree-level diagrams with these mediators are shown in Fig.~\ref{fig:diagram}.

Most models explaining these anomalies have so far relied on the left-handed (LH) SM neutrinos to provide the missing energy in the $\bar{B} \rightarrow D^{(*)} \tau \nu$ decays. However, recently there has been increased interest in the possibility that right-handed (RH) sterile neutrinos are instead present in the decays \cite{He:2012zp,Dutta:2013qaa,Cline:2015lqp,Becirevic:2016yqi,Bardhan:2016uhr,Dutta:2016eml,Iguro:2018qzf,Asadi:2018wea,Greljo:2018ogz,Abdullah:2018ets,Azatov:2018kzb,Heeck:2018ntp,Carena:2018cow,Iguro:2018fni}. Specifically, it was shown in \cite{Asadi:2018wea,Greljo:2018ogz} that a $W'$ coupling to light RH neutrinos could explain both anomalies, while evading severe bounds from flavor physics and direct collider searches that rule out $W'$ models with LH neutrinos \cite{Greljo:2015mma,Faroughy:2016osc}.

With the large number of mediators proposed to explain the $R_{D^{(*)}}$ anomalies, it is important to understand which are phenomenologically viable, and to figure out ways to distinguish them  experimentally \cite{Bardhan:2016uhr,Dutta:2016eml,Duraisamy:2013kcw,Becirevic:2016hea,Alonso:2016gym,Altmannshofer:2017poe,Alok:2017qsi,Alok:2018uft,Huang:2018nnq}. In particular, it is interesting to ask: what measurements can we make in order to tell the difference between models with LH and RH neutrinos?
In this paper, we will explore the possibility of using various angular and polarization asymmetry observables for this purpose.  Of particular interest here are the forward-backward asymmetry of the leptonic pair in both $\bar{B}\rightarrow D\tau \nu$ and $\bar{B}\rightarrow D^*\tau \nu$ decays \cite{Bardhan:2016uhr,Duraisamy:2013kcw,Becirevic:2016hea,Alonso:2016gym,Sakaki:2012ft,Datta:2012qk,Ivanov:2015tru,Alok:2016qyh,Ivanov:2017mrj,Alonso:2017ktd} and the asymmetry in the polarization of the $\tau$ lepton in the decays \cite{Fajfer:2012vx,Bardhan:2016uhr,Becirevic:2016hea,Alonso:2016gym,Ivanov:2017mrj,Alonso:2017ktd,Tanaka:1994ay,Tanaka:2010se}.\footnote{Another commonly studied observable is the differential decay rate ${d\Gamma\over dq^2}$, see for example \cite{Lees:2013uzd,Bardhan:2016uhr,Tanaka:2012nw}. We find that this observable is less useful for distinguishing between different models with different types of neutrinos; see App.~\ref{sec:analytics} for a discussion, in particular Fig.~\ref{fig:q2dist}.  }

We will focus on the future measurement of these asymmetry observables at Belle~II.\footnote{While LHCb can also provide us with an enormous dataset, due to the large background in this hadronic collider and reduced kinematic data (e.g. lack of knowledge of the initial rest-frame of the $B$ mesons), it can be limited in some precision measurements.}
An upgrade of the Belle experiment, Belle~II is an $e^+e^-$ collider with asymmetric beams and center of mass energy of $\sim10 $~GeV, producing $\Upsilon (4S)$ which subsequently decay to pairs of $B$ mesons.
It is projected to collect more than forty times the data of Belle (around  $50~\mathrm{ab}^{-1}$, a total of $\sim 55$ million $B\bar{B}$ pairs) by 2025. 
With the total planned dataset, the uncertainty on $R_D$ ($R_{D^*}$) is expected to be as low as $\sim 3\%$ ($\sim 2\%$) \cite{Kou:2018nap}. If the current global average (\ref{eq:rdobs}) persists after Belle~II, it will indicate an undisputed discovery of new physics. In the case of such a discovery, the asymmetry observables we study in this paper can provide information about the beyond-the-SM physics responsible for these anomalies. 

In this work we calculate the contribution to these observables for all the different types of proposed mediators, and we present our results as numerical formulas for each observable in terms of the Wilson coefficients of the general dimension-6 effective Hamiltonian. Previous studies \cite{Bardhan:2016uhr,Dutta:2016eml,Duraisamy:2013kcw,Becirevic:2016hea,Alonso:2016gym,Hu:2018veh} have considered how to separate different explanations of the $R_{D^{(*)}}$ anomalies using  angular observables. In this paper, we include additional operators involving new RH neutrinos in the comparison and specifically aim to distinguish simplified models with different neutrino chiralities. We also take the experimental limitations into account and focus on the observables for which there are proposed measurement strategies.

Besides requiring that these models simultaneously explain both the $R_D$ and $R_{D^*}$ anomalies, we also impose other experimental constraints (in particular, modifying the branching ratio for the $B_c\rightarrow \tau \nu$ \cite{Li:2016vvp,Alonso:2016oyd,Celis:2016azn, Akeroyd:2017mhr,Azatov:2018knx} and $b\rightarrow s \nu \nu$ processes \cite{Grossman:1995gt,Altmannshofer:2009ma,Sakaki:2013bfa,Dumont:2016xpj,Kumar:2018kmr}). With all the experimental constraints taken into account, the list of currently viable mediators that can individually explain the anomalies can be found in Tab.~\ref{tab:viables}. (We use the nomenclature from \cite{Dorsner:2016wpm} for the LQs.) 
We show that these asymmetry observables can significantly differentiate between models with LH or RH neutrinos, even after taking into account projections of future experimental precision  \cite{Alonso:2017ktd}. 

\begin{table}
\begin{tabular}{|c|c|c|c|}
\hline 
 Mediator & SM Charges  &Type of Neutrinos \\ 
\hline 
\hline 
$S_1$ scalar LQ & $(\bar{3},1,1/3)$ & LH or RH \\ 
\hline 
$U_1$ vector LQ & $(3,1,2/3)$  & LH or RH  \\ 
\hline 
$R_2$ scalar LQ & $(3,2,7/6)$ &  LH   \\ 
\hline 
$W'$ & $(1,1,1)$ & RH  \\
\hline 
\end{tabular} 
\caption{ The list of currently viable single mediator solutions to the anomalies and their charges under the SM gauge groups $\left(SU(3), SU(2), U(1) \right)$. We further indicate the type of neutrino they require in the $\bar{B}$ meson decay to explain the anomalies. }
\label{tab:viables}
\end{table}

The questions of what mediators remain viable and how distinguishable they are from one another after Belle~II depend heavily on the $R_{D^{(*)}}$ ratios measured by the new experiment. To highlight the power of these asymmetry observables, in this work we will consider two hypothetical outcomes for the global averages after Belle~II, which have substantially different implications for our study. In the first scenario, we imagine that Belle~II will measure  $R_{D^{(*)}}$ equal to their current global averages. (This would correspond to a $\sim 10\sigma$ discrepancy with the SM.) In the second scenario, we posit that  $R_{D^{(*)}}$ will be reduced, but remain 5$\sigma$ discrepant with the SM. As we will show, in the 10$\sigma$ scenario, not only the neutrino chiralities can be easily distinguished from one another using the asymmetry observables, but even individual models with the same neutrino chiralities can be told apart. In the 5$\sigma$ scenario, we show that the neutrino chiralities can still always be distinguished from each other; in order to distinguish a small subset of models we require additional CP-odd polarization asymmetry measurements. A measurement strategy for these CP-odd observables is yet to be delivered, but due to their discriminating power, they should be considered a high priority. Our results highlight the importance of finding experimental strategies for their measurement.

The outline of our paper is as follows. In Sec.~\ref{sec:models}, we describe those single operators and simplified models which can explain both $R_{D^{(*)}}$ anomalies and survive other experimental constraints. We also define in detail the two different scenarios for Belle~II measurements of $R_{D^{(*)}}$ described above, and show how these measurements alone can significantly reduce the set of viable models. 
In Sec.~\ref{sec:obsevables}, we define the angular observables (forward-backward and polarization asymmetries) and calculate their dependence on the Wilson coefficients. We further discuss their experimental status and review a recent proposal   \cite{Alonso:2017ktd} with higher projected sensitivity for their measurements at Belle~II. Finally, in Sec.~\ref{sec:discerning}, we show which combinations of angular observables can be used to distinguish the viable models with LH and RH neutrinos. We show that for different outcomes at Belle~II, we will be able to tell different types of neutrinos apart, and in almost all cases can distinguish individual models as well. 
 We conclude in Sec.~\ref{sec:conclusion} with a brief summary and outlook.

Several appendices are included in the end. In App.~\ref{sec:hadronicfuncs} we list the leptonic matrix elements used in our calculation as well as some hadronic functions needed for the calculation involving RH neutrinos. App.~\ref{sec:analytics} includes further details on the calculation of the asymmetries and full analytic formulas for each of them. Finally, in App.~\ref{sec:scan} we point out a linear relationship between different CP-even observables we study in this work and explain a numerical scan that we perform over the viable range of Wilson coefficients.

\section{Simplified Models for $R_{D^{(*)}}$}
\label{sec:models}

The set of all possible dimension-6 operators modifying the $b\to c\tau\nu$ decay rate can be written as
\beq\label{Heff}
 {\mathcal H}_{\rm eff} = \frac{4 G_F V_{cb}}{\sqrt{2}}  \left( {\mathcal O}^V_{LL} + \sum_{X=S,V,T\atop M,N=L,R} C^X_{MN}{\mathcal O}^X_{MN} \right)
\eeq
where the pre-factor normalizes the SM Wilson coefficient to unity, and the four-fermion effective operators are defined as
\begin{eqnarray}
 {\mathcal O}^S_{MN} & \equiv & (\bar c P_M b)(\bar \tau P_N \nu) \nonumber \\
{\mathcal O}^V_{MN} & \equiv &(\bar c \gamma^\mu P_M b)(\bar \tau \gamma_\mu P_N \nu) \label{eq:Lop}\\
{\mathcal O}^T_{MN} &\equiv &(\bar c \sigma^{\mu\nu} P_M b)(\bar \tau \sigma_{\mu\nu}P_N \nu), \nonumber
\end{eqnarray}
for $M,N = R$ or $L$. These operators can be generated by integrating out heavy new mediators; the Wilson coefficients $C^X_{MN}$ parametrize the most general contribution.\footnote{The tensor operators with $M \neq N$, $\mathcal{O}^T_{RL}$ and $\mathcal{O}^T_{LR}$, are identically zero. To generate $\mathcal{O}^V_{LR}$ and $\mathcal{O}^V_{RL}$ gauge-invariantly, we further need to insert some Higgs field vacuum expectation values. They can be absorbed into the Wilson coefficients. } Different UV models can be categorized using the operators they give rise to (typically more than one), see Sec.~\ref{subsec:MM}.

In the operator basis of \eqref{Heff}, the contribution of new physics to the ratios $R_{D^{(*)}}$ can be calculated in terms of the ten (possibly complex) Wilson coefficients: five involving a SM left-handed neutrino, and five requiring a new right-handed neutrino. The numerical contribution of all the operators from \eqref{Heff} to the ratios are \cite{Asadi:2018wea}:
\bea
\label{eq:allCRD}
R_D &  \approx   R_D^{SM} \times \left\lbrace \left(  |1+C^V_{LL}+C^V_{RL}|^2 + |C^V_{RR}+C^V_{LR}|^2 \right)    \right.  \\
&  +  1.35  \left( |C^S_{RL}+C^S_{LL}|^2 + |C^S_{LR}+C^S_{RR}|^2 \right)  + 0.70 	\left( |C^T_{LL}|^2 +  |C^T_{RR}|^2 \right)    \\
& +   1.72 \mathcal{R}e \left[ (1+C^V_{LL}+C^V_{RL})(C^S_{RL}+C^S_{LL})^* + (C^V_{RR}+C^V_{LR})(C^S_{LR}+C^S_{RR})^* \right]     \\
& + \left. 1.00  \mathcal{R}e \left[  (1+C^V_{LL}+C^V_{RL})(C^T_{LL})^*  +   (C^V_{LR}+C^V_{RR})(C^T_{RR})^* \right] \right\rbrace   ,\\\\
R_{D^{*}} & \approx  R_{D^*}^{SM}\times  \left\lbrace  \left( |1+C^V_{LL}|^2+|C^V_{RL}|^2	+ |C^V_{LR}|^2+|C^V_{RR}|^2	\right)  \right.  \\
& +  0.04	\left( |C^S_{RL}-C^S_{LL}|^2 +  |C^S_{LR}-C^S_{RR}|^2 \right)   \\
& +   12.11 \left(  |C^T_{LL}|^2 +  |C^T_{RR}|^2 \right)   - 1.78  \mathcal{R}e \left[ (1+C^V_{LL})(C^V_{RL})^* +  C^V_{RR} (C^V_{LR})^* \right]  \\
& + 5.71 \mathcal{R}e \left[	C^V_{RL} (C^T_{LL})^*	+  C^V_{LR} (C^T_{RR})^*	\right]   -  4.15				\mathcal{R}e \left[	(1+C^V_{LL}) (C^T_{LL})^*	+ C^V_{RR} (C^T_{RR})^*	\right]   \\
& +  \left. 0.12 \mathcal{R}e\left[	(1+C^V_{LL}-C^V_{RL}) (C^S_{RL}-C^S_{LL})^*	+  (C^V_{RR}-C^V_{LR}) (C^S_{LR}-C^S_{RR})^*\right] \right\rbrace  . 
\eea
\medskip
Further details on deriving these numerical equations are included in App.~\ref{sec:analytics}.

\subsection{Single Operator Solutions}
\label{subsec:single operators}

\begin{figure}
\includegraphics[scale=0.7]{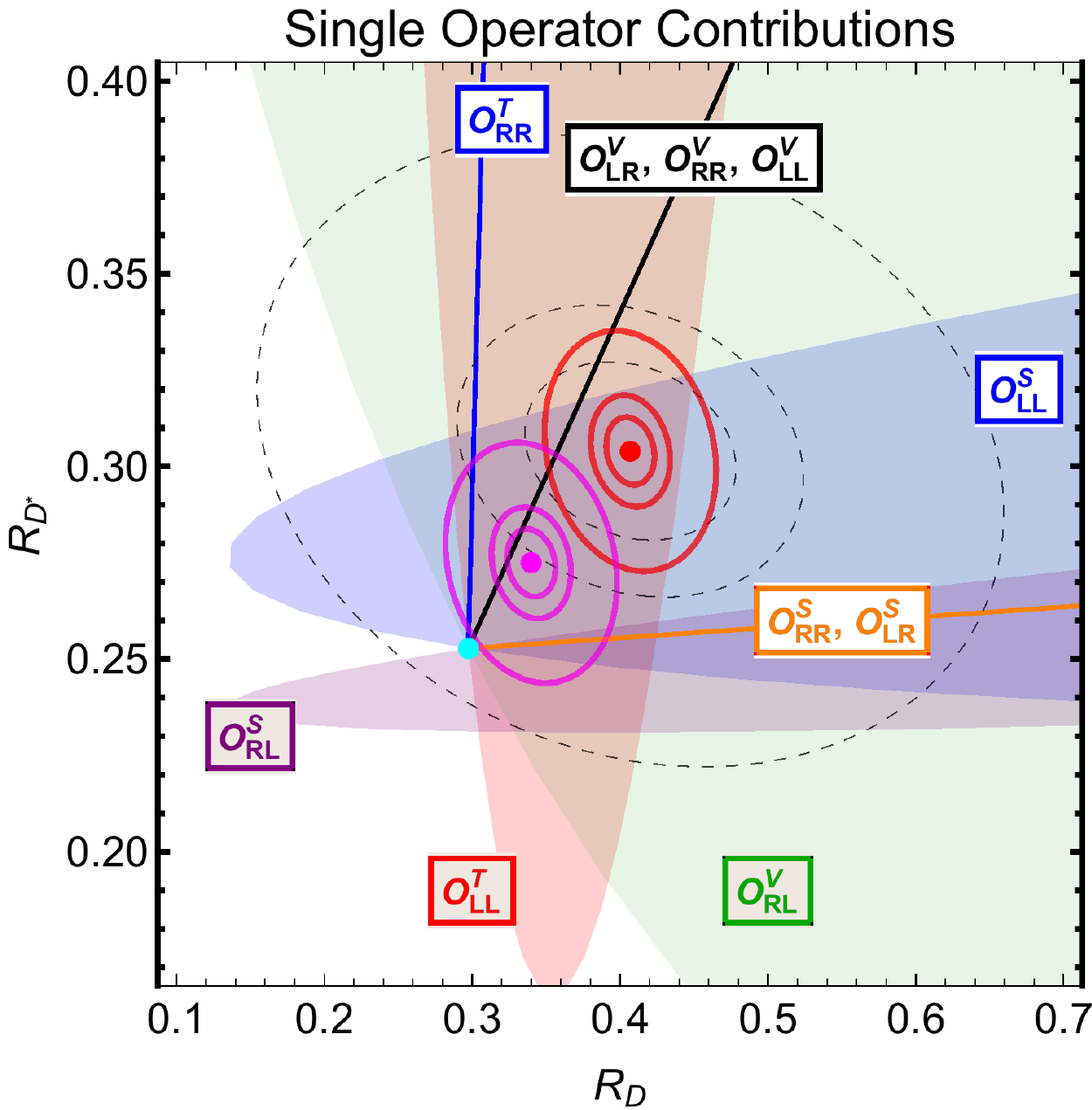}
\caption{Ranges of $R_{D^{(*)}}$ spanned by single operators with complex Wilson coefficients. The SM prediction is denoted by a cyan dot. No other experimental constraints are imposed in this figure. The 1, 2, and $5\sigma$ contours around the current global average are shown as gray-dashed lines. We also show these contours with the projected Belle~II precision \cite{Kou:2018nap} around the current global average (red ellipses) and a hypothetical average after Belle~II that still barely allows a $5\sigma$ discovery (magenta ellipses), assuming the current correlation $\rho_{\rm corr}=-0.2$. (See Sec.~\ref{subsec:scenarios} for details.) }
\label{fig:rdrdsrange_operators}
\end{figure}

The range of $R_{D^{(*)}}$ that each individual operator can generate (with general complex Wilson coefficients) is indicated in Fig.~\ref{fig:rdrdsrange_operators}, along with the present-day experimental and theoretical combined uncertainty in the $R_{D^{(*)}}$ measurements, showing the 1, 2, and $5\sigma$ contours (gray-dashed ellipses). For a review of experimental correlations in the measurements of $R_{D^{(*)}}$, see \cite{Lees:2013uzd,Huschle:2015rga}. In Fig.~\ref{fig:rdrdsrange_operators}, 
we use the current average of the correlations, $\rho_{\rm corr} = -0.2$ \cite{HFLAV16}. We see that out of all ten effective operators in \eqref{Heff}, there are only six that can explain both anomalies simultaneously: $\mathcal{O}^V_{LL}$, $\mathcal{O}^V_{RL}$, $\mathcal{O}^S_{LL}$, $\mathcal{O}^T_{LL}$, $\mathcal{O}^V_{RR}$, and $\mathcal{O}^V_{LR}$. 

We are not aware of any UV-complete models in the literature for these anomalies that rely solely on any of the operators $\mathcal{O}^V_{RL}$, $\mathcal{O}^T_{LL}$, or $\mathcal{O}^V_{LR}$. Despite this lack of UV-complete models, we will include these three single operator explanations in our analysis for the sake of completeness.

\subsection{Simplified Model Solutions}
\label{subsec:MM}

We can now enumerate the full set of ``simplified" models that can explain both the $R_{D^{(*)}}$ anomalies. In this context, ``simplified" means a single new mediator particle that can be integrated out to provide one or more of the effective operators which modify $R_{D^{(*)}}$. 

An over-complete list of all the simplified models that can generate the operators in \eqref{Heff} with LH or RH neutrinos can be found in \cite{Sakaki:2013bfa,Freytsis:2015qca,Robinson:2018gza}. We gather these mediators in Tab.~\ref{tab:operators}. Notice that the $S_1$ and $U_1$ LQs and uncolored mediators can couple to either LH or RH fermions and so give rise to operators involving either type of neutrinos. In this work we consider these possibilities as separate solutions to the anomalies and will try to distinguish them from one another. 

The factor of $x$ in Tab.~\ref{tab:operators} relates the Wilson coefficients of scalar and tensor operators in some models after Fierz transformation. At the mediator scale, $x=1/4$ for all the models in Tab.~\ref{tab:operators}; as we run down to the GeV scale $x$ changes to $\sim 1/8$ \cite{Dorsner:2013tla,Sakaki:2014sea,Angelescu:2018tyl}, with the exact value depending on the mediator scale. For simplicity, we use the fiducial value $x=1/8$ in our analysis.

\begin{table}
\begin{tabular}{|c|c|c|}
\hline 
\small{Mediator} & \small{Operator Combination} & Viability \\ 
\hline 
\hline 
Colorless Scalars  &\small{$\mathcal{O}^S_{XL}$} & $\xmark$ ($Br \left(B_c \rightarrow \tau \nu \right)$)  \\
\hline 
$W'^\mu$ (LH fermions) &\small{$\mathcal{O}^V_{LL}$} & $\xmark$ (collider bounds)  \\
\hline 
\small{$S_1$ LQ $(\bar{3},1,1/3)$ (LH fermions)  } & \small{$\mathcal{O}^S_{LL}- x \mathcal{O}^T_{LL}, \,\,\,  \textcolor{red}{\mathcal{O}^V_{LL}}$} &  $\cmark$ \\
\hline
\small{$U_1^\mu$ LQ $(3,1,2/3)$ (LH fermions) }&\small{$\mathcal{O}^S_{RL},\,\,\,  \mathcal{O}^V_{LL} $} & $\cmark$ \\
\hline
\small{$R_2$ LQ $(3,2,7/6)$ }&\small{$\mathcal{O}^S_{LL}+ x \mathcal{O}^T_{LL} $} & $\cmark$ \\
\hline
\small{$S_3$ LQ $(\bar{3},3,1/3)$}  &\small{$ \mathcal{O}^V_{LL} $} & $\xmark$ ($b\rightarrow s \nu\nu $) \\
\hline
\small{$U_3^\mu$ LQ $(3,3,2/3)$}  &\small{$ \mathcal{O}^V_{LL} $} & $\xmark$ ($b\rightarrow s \nu\nu $) \\
\hline
\small{$V_2^\mu$ LQ $(\bar{3},2,5/6)$ }&\small{$\mathcal{O}^S_{RL}$} & $\xmark$ ($R_{D^{(*)}}$ value) \\
\hline
\hline
Colorless Scalars &\small{$\mathcal{O}^S_{XR}$} & $\xmark$ ($Br \left(B_c \rightarrow \tau \nu \right)$)  \\
\hline
$W'^\mu$ (RH fermions) &\small{$\mathcal{O}^V_{RR}$} & $\cmark$  \\
\hline
\small{$\tilde{R}_2$ LQ $(3,2,1/6)$  }&\small{$\mathcal{O}^S_{RR}+ x \mathcal{O}^T_{RR} $} & $\xmark$ ($b\rightarrow s \nu\nu $) \\
\hline
\small{$S_1$ LQ $(\bar{3},1,1/3)$ (RH fermions)  }&\small{$\mathcal{O}^V_{RR},\,\,\, \textcolor{red}{\mathcal{O}^S_{RR}- x \mathcal{O}^T_{RR}}$}  & $\cmark$  \\
\hline
\small{$U_1^\mu$ LQ $(3,1,2/3)$ (RH fermions)  }&\small{$\mathcal{O}^S_{LR},\,\,\,   \mathcal{O}^V_{RR} $}  & $\cmark$  \\
\hline
\end{tabular} 
\caption{ A complete list of the simplified mediator models and resulting effective operators that are possibly relevant for the $R_{D^{(*)}}$ anomalies. The $U_1^\mu$ and $S_1$ LQs as well as the colorless scalars can give rise to two independent Wilson coefficients, while the rest of the mediators can generate only one. We use $x=1/8$ in this work, see the text for more details. We indicate in the last column if the model is still viable (by $\cmark$) and if not, what experimental constraint rules it out (see Sec.~\ref{subsec:constraints} for discussion of these constraints). The operators in red are severely constrained by the $b\rightarrow s\nu\nu$ constraints as well.
}
\label{tab:operators}
\end{table}

\begin{figure}
\includegraphics[scale=0.55]{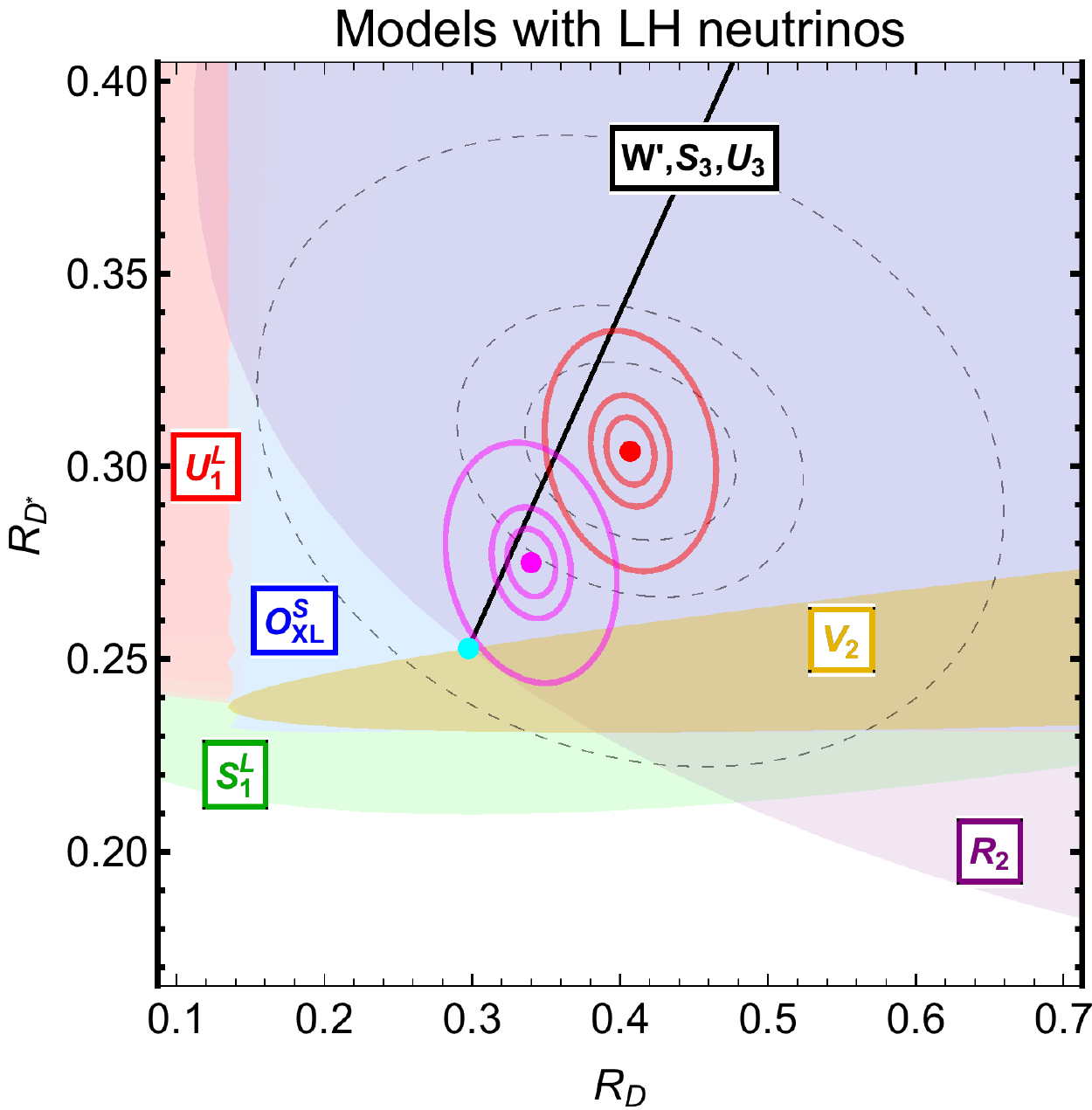}
\includegraphics[scale=0.55]{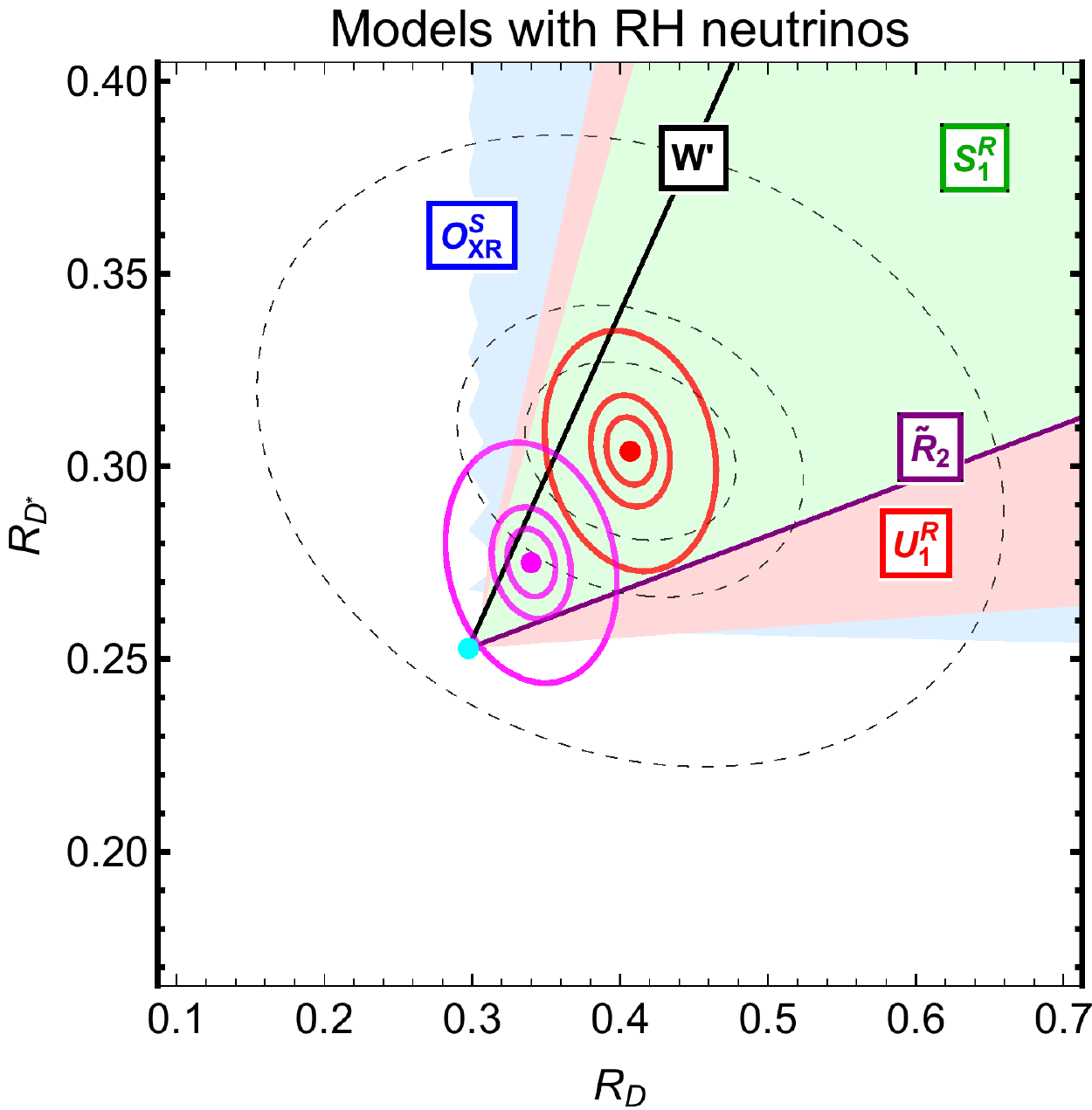}
\caption{The range of  $R_{D^{(*)}}$ spanned by the simplified models from Tab.~\ref{tab:operators} with complex Wilson coefficients. 
 The superscript on $S_1$ and $U_1$ LQ refers to the neutrino chirality which they are coupled to in each figure. No other experimental constraints are imposed in this figure. The other features are as in Fig.~\ref{fig:rdrdsrange_operators}.
 }
\label{fig:rdrdsrange_models}
\end{figure}

In Fig.~\ref{fig:rdrdsrange_models}, we show the values of $R_D$ and $R_{D^*}$ which can be obtained by each of the relevant mediators in Tab.~\ref{tab:operators}, scanning over complex Wilson coefficient(s). In these plots the superscripts $L$ and $R$ on $S_1$ and $U_1$ LQs refer to the neutrino chirality they couple to. Some mediators yield lines in this parameter space; these are single-coefficient models whose contribution to $R_D$ and $R_{D^*}$ are independent of the phase of the coefficient. Other operators can cover a region of $R_{D^{(*)}}$ as the coefficients are varied, either because the $R_{D^{(*)}}$ values depend on both magnitude and phase of single operator, or the model results in two independent Wilson coefficients.

\subsection{Additional Constraints and Final List of Viable Models}
\label{subsec:constraints}

In addition to explaining $R_{D^{(*)}}$, a viable mediator must also avoid a number of other stringent constraints. In this subsection we will review these and then list the surviving viable solutions. 

A subset of the couplings which modify the $\bar{B}\rightarrow  D^{(*)}\tau \nu$ decay can enhance the branching ratio $B_c\rightarrow  \tau \nu$ \cite{Li:2016vvp,Alonso:2016oyd,Celis:2016azn, Akeroyd:2017mhr,Azatov:2018knx}. In terms of the Wilson coefficients in \eqref{Heff},
\begin{eqnarray}
\frac{Br (B_c \rightarrow \tau \nu )}{Br (B_c \rightarrow \tau \nu )|_{\rm SM}} &=& \left|1+ \left( C^V_{LL}  - C^V_{RL}\right)  +   \frac{m_{B_c}^2}{m_\tau (m_b+m_c)} \left( C^S_{RL}  - C^S_{LL}\right)  \right|^2 \nonumber \\
\label{eq:Bcratio}
&+& \left|\left( C^V_{RR}  - C^V_{LR}\right)  +   \frac{m_{B_c}^2}{m_\tau (m_b+m_c)} \left( C^S_{LR}  - C^S_{RR}\right)  \right|^2.
\end{eqnarray}
Given the mass ratios above, these equations imply tighter bounds on the scalar operators than the vector ones. The SM prediction is $Br (B_c \rightarrow \tau \nu )|_{\rm SM} \sim 2\%$. The $B_u \rightarrow \tau \nu$ decay in LEP at the $Z$ boson peak can be used to place the constraint \cite{Akeroyd:2017mhr}
\begin{equation}
Br (B_c \rightarrow \tau \nu ) \leqslant 10 \% ,
\label{eq:Bcbound}
\end{equation}
which in turn puts a constraint on the possible Wilson coefficients in \eqref{eq:Bcratio}. Using the theoretical calculation of the $B_c$ lifetime and its uncertainties, a looser bound of $Br (B_c \rightarrow \tau \nu ) \leqslant 30 \% $ can be obtained as well \cite{Alonso:2016oyd}. These branching ratio constraints put particularly severe bounds on models relying on $\mathcal{O}^S_{MN}$ operators to explain the anomalies -- to the extent that if a model relies solely on a scalar operator to explain the anomalies, it is ruled out by the constraint (\ref{eq:Bcbound}). This remains true even if the global average of the anomalies reduces to the magenta dot in Fig.~\ref{fig:rdrdsrange_models} after Belle~II.

The other relevant flavor constraint is from $b\rightarrow s \nu\nu $ decay and the meson decays it enables \cite{Grossman:1995gt,Altmannshofer:2009ma,Sakaki:2013bfa}, in particular the inclusive $B\rightarrow X_s \nu \nu$ and the exlusive $B\rightarrow K^{(*)} \nu \nu$. The current bound on the inclusive branching ratio of $B\rightarrow X_s \nu \nu$ is from the ALEPH Collaboration \cite{Barate:2000rc}, 
\begin{equation}
Br \left(	B\rightarrow X_s \nu \nu	\right) \leqslant 6.4 \times 10^{-4} 
\label{eq:bsnuboundincl}
\end{equation}
at $90\%$ CL, whereas the bound on the exclusive decay rates above are \cite{Grygier:2017tzo}
\begin{equation}
Br \left(	B\rightarrow K \nu \nu	\right) \leqslant 1.6 \times 10^{-5}, ~~~ Br \left(	B\rightarrow K^* \nu \nu	\right) \leqslant 2.7 \times 10^{-5}.
\label{eq:bsnuboundexcl}
\end{equation}

While the mediators introduced for $R_{D^{(*)}}$ generate charged currents, the $b\rightarrow s \nu \nu$ decay requires a neutral current beyond the SM. However, in some models that rely on leptoquarks \cite{Sakaki:2013bfa,Dumont:2016xpj,Robinson:2018gza}, there is an inevitable neutral current due to the SM $SU(2)_L$ symmetry. 

If both the neutrinos in the $b\rightarrow s \nu \nu$ decay are LH, Lorentz invariance implies that the dimension six effective operator can only be a vector current.
The associated charged current then can only give rise to $\mathcal{O}^V_{LL}$. Thus, for the models with LH neutrinos, this bound may only constrain the $C^V_{LL}$ Wilson coefficient.

For instance, the $S_3$ LQ can give rise to the following terms (among others) \cite{Dumont:2016xpj} 
\beq
\mathcal{L} \supset  g_L^{ij} \bar{Q}_L^{c,i} i \sigma_2 \sigma^a L_L^j S_3^a,
\label{eq:LQsbsnunu}
\eeq
where $i,j$ are flavor indices and $a$ is an $SU(2)$ adjoint index. After Fierz transformation, this LQ can give rise to $\mathcal{O}^V_{LL}$ with
\begin{equation}
C^V_{LL} = - \frac{V_{tb}}{V_{cb}} \frac{g_L^{3j_1} g_L^{23,*}}{4\sqrt{2} G_F M^2_{S_3}},
\label{eq:bsnuCVLL}
\end{equation}
where $G_F$ is the fermi constant and $M_{S_3}$ is the $S_3$ LQ mass. Due to the $SU(2)_L$ symmetry, this term will contribute to $b\rightarrow s \nu \nu$ as well.\footnote{It is possible to generate $C^V_{LL}$ with these leptoquarks by invoking $g^{i\neq3,j}_L$ couplings as well. In this case, however, we will have a substantial CKM suppression and will need non-perturbative couplings to explain the anomalies. As a result, we discard this possibility.} The contribution of this LQ to the neutral $b\rightarrow s \nu \nu$ processes can be captured by the following effective Hamiltonian \cite{Altmannshofer:2009ma}
\begin{equation}
\mathcal{H}_{\mathrm{eff}} \supset - \sqrt{2} \frac{\alpha_{em}}{\pi} G_F V^{*}_{ts} V_{tb} C^\nu_{L} \left(		\bar{s} \gamma^\mu P_L	b \right) \left(		\bar{\nu} \gamma_\mu 	P_L \nu \right),
\label{eq:Heffbsnunu}
\end{equation}
where $C^\nu_L$ is a Wilson coefficient and $\alpha_{em}$ is the fine structure constant. After integrating out a $S_3$ LQ, the generated $C^\nu_L$ Wilson coefficient will be
\begin{equation}
C^\nu_L = \frac{\pi}{2\sqrt{2} \alpha_{em}G_F V^{*}_{ts}V_{tb}} \frac{g^{2j_1}_Lg^{3j_2,*}_{L}}{M_{S_3}^2},
\label{eq:CvLS3}
\end{equation}
where $j$ indices refer to different generations of neutrinos. Using the numerical formulas reported in \cite{Altmannshofer:2009ma} and the bound on $Br \left(	B\rightarrow K \nu \nu	\right)$, which is the most constrained branching ratio in \eqref{eq:bsnuboundincl}-\eqref{eq:bsnuboundexcl}, we find
\begin{equation}
|g_L^{3j_1} g_L^{2j_2,*}| \frac{1 \mathrm{TeV}^2}{M_{S_3}^2} \lesssim 0.017,
\label{eq:watanabebsnunu}
\end{equation}
which when combined with \eqref{eq:bsnuCVLL} yields:
\begin{equation}
C^V_{LL} \lesssim 0.006.
\label{eq:bsnunu}
\end{equation}

This bound is severe enough that we can safely neglect the contribution of $C^V_{LL}$ from the $S_3$ LQ to the anomalies. A similar bound also applies to the $U_3$ and $S_1$ LQs that are coupled to LH fermions \cite{Dumont:2016xpj}. $S_3$ and $U_3$ can only generate $\mathcal{O}^V_{LL}$ and are therefore completely ruled out. 
Since $S_1$ can generate $\mathcal{O}^S_{LL}$ and $\mathcal{O}^T_{LL}$ operators from other couplings in the Lagrangian, it can still be a viable explanation of the anomalies despite this severe bound on $C^V_{LL}$. Finally, due to the $SU(2)$ structure of the operators that it gives rise to, this bound does not apply to $U_1$ LQ \cite{Grossman:1995gt,Dumont:2016xpj}, even though this LQ does generate $\mathcal{O}^V_{LL}$. 

If instead we allow for one of the neutrinos in the $b\rightarrow s\nu\nu$ process to be RH, then the dimension six effective operator can be either a scalar or a tensor current. In particular, the same couplings that generate $\mathcal{O}^S_{RR}\pm x \mathcal{O}^T_{RR}$ operators in $S_1$ and $\tilde{R}_2$ LQs also give rise to the operators \cite{Robinson:2018gza}
\begin{equation}
\left(	\bar{s}_L b_R	\right)		\left(		\bar{\nu}_L \nu_R 	\right),~~~ \left(	\bar{s}_L \sigma^{\mu\nu} b_R	\right)		\left(		\bar{\nu}_L \sigma_{\mu\nu} \nu_R 	\right),
\label{eq:bsnunuRHops}
\end{equation}
which contribute to the $b\rightarrow s\nu\nu$ processes. The bound on these operators Wilson coefficients translates into $\mathcal{O}(0.01)$ bounds on the $\mathcal{O}^S_{RR}$ in $S_1$ and $\tilde{R}_2$ models \cite{Robinson:2018gza}, hence we can safely discard their contribution to the anomalies too.\footnote{Notice that since these models do not have any interference with the SM, the contribution to the anomalies is quadratic in their Wilson coefficient and a $\mathcal{O}(0.01)$ bound on a Wilson coefficient implies order $10^{-4}$ improvement in the $R_{D^{(*)}}$ ratios. } The $\tilde{R}_2$ is thus ruled out, while the $S_1$ LQ model becomes degenerate with a $W'$ and the single operator $C^V_{RR}$.

Other than these flavor constraints, there are some bounds from direct searches for these mediators. For the case of leptoquarks, the current bounds are not severe enough to rule out any further models \cite{Robinson:2018gza,Angelescu:2018tyl,Monteux:2018ufc}. On the other hand, the bounds on the $W'$ are fairly constraining \cite{Asadi:2018wea,Greljo:2018ogz,Faroughy:2016osc,Kumar:2018kmr}. In particular, if the $W'$ couples to LH fermions, the bounds on the accompanying $Z'$ effectively rule out the explanations of the anomalies \cite{Faroughy:2016osc,Kumar:2018kmr}. 

The combination of these constraints significantly reduces the viable explanations of the $R_{D^{(*)}}$ anomalies. In the last column of Tab.~\ref{tab:operators} we indicate which models survive. In all, there are three viable simplified models ($S_1$, $R_2$, and $U_1$ LQs) that couple to LH neutrinos, and three that couple to RH ($W'$, $U_1$ and $S_1$ LQs). Note however that the $W'$ and $S_1$ LQ with RH neutrinos generate the same Wilson coefficient, and this a subset of the parameter space generated by the $U_1$ LQ with RH neutrinos.  In the rest of this paper, we will focus on these surviving simplified models, along with the viable single operators $\mathcal{O}^V_{LR}$, $\mathcal{O}^V_{RL}$, and $\mathcal{O}^T_{LL}$.

\subsection{Benchmark Belle~II Scenarios}
\label{subsec:scenarios}

Belle~II will measure $R_{D^{(*)}}$ with much smaller errors compared to the present, thus greatly reducing the possible range of Wilson coefficients in each model. 
As can be seen in Figs.~\ref{fig:rdrdsrange_operators}--\ref{fig:rdrdsrange_models}, central values near the present averages would by themselves rule out at high significance many models which are presently under consideration. Meanwhile, values closer to the SM prediction (while still allowing a $5\sigma$ discovery at Belle~II) would leave all the mediators and single operators we currently consider as possibilities, before constraints from the asymmetry observables are applied. Aside from having a potentially huge impact on the list of models that explain the anomalies, this can also greatly affect our ability to distinguish between these models with further measurements (such as the asymmetries).

As a result, we will consider two different outcomes of the Belle~II measurement of $R_{D^{(*)}}$ as benchmarks for our study. 
\begin{enumerate}
\item {\it The 10$\sigma$ scenario:} Belle~II measures $R_{D^{(*)}}$ with central values equal to the present average. With the projected Belle~II sensitivities, this would correspond to a $\mathcal{O}\left( 10\sigma \right)$ discovery. We then consider ranges of $R_{D^{(*)}}$ within the $2\sigma$ Belle~II error ellipse about this central value (the second innermost red ellipse in Fig.~\ref{fig:rdrdsrange_operators}--\ref{fig:rdrdsrange_models}). As we will show, in the 10$\sigma$ scenario, the task of discerning different models is simplified considerably.

\item {\it The 5$\sigma$ scenario:} The measured $R_{D^{(*)}}$ values are closer to the SM expectation while still allowing a $5\sigma$ discovery; specifically, we assume the central value of the anomalies after Belle~II shifts to $R_D = 0.34$ and $R_{D^*}=0.275$. This point was chosen to have $5\sigma$ significance with Belle~II projected error bars, to be within $\sim 2\sigma$ of the current global average, and (crucially) to allow for all of the simplified models to continue to explain the $R_{D^{(*)}}$ anomalies (see Fig.~\ref{fig:rdrdsrange_models}). Compared to the 10$\sigma$ scenario, distinguishing between different models is much more challenging here. 
\end{enumerate}

These two benchmark scenarios are meant to bracket the range of possibilities that we can expect from Belle~II, assuming that the $R_{D^{(*)}}$ anomalies are fully confirmed. The $10\sigma$ scenario is meant to illustrate how easy it can be to distinguish different models using the $\tau$ asymmetries, while the $5\sigma$ scenario is meant to provide a ``worst-case scenario" from the point of view of distinguishing between different models.

\section{Asymmetry Observables}
\label{sec:obsevables}

\begin{figure}
\includegraphics[scale=0.5]{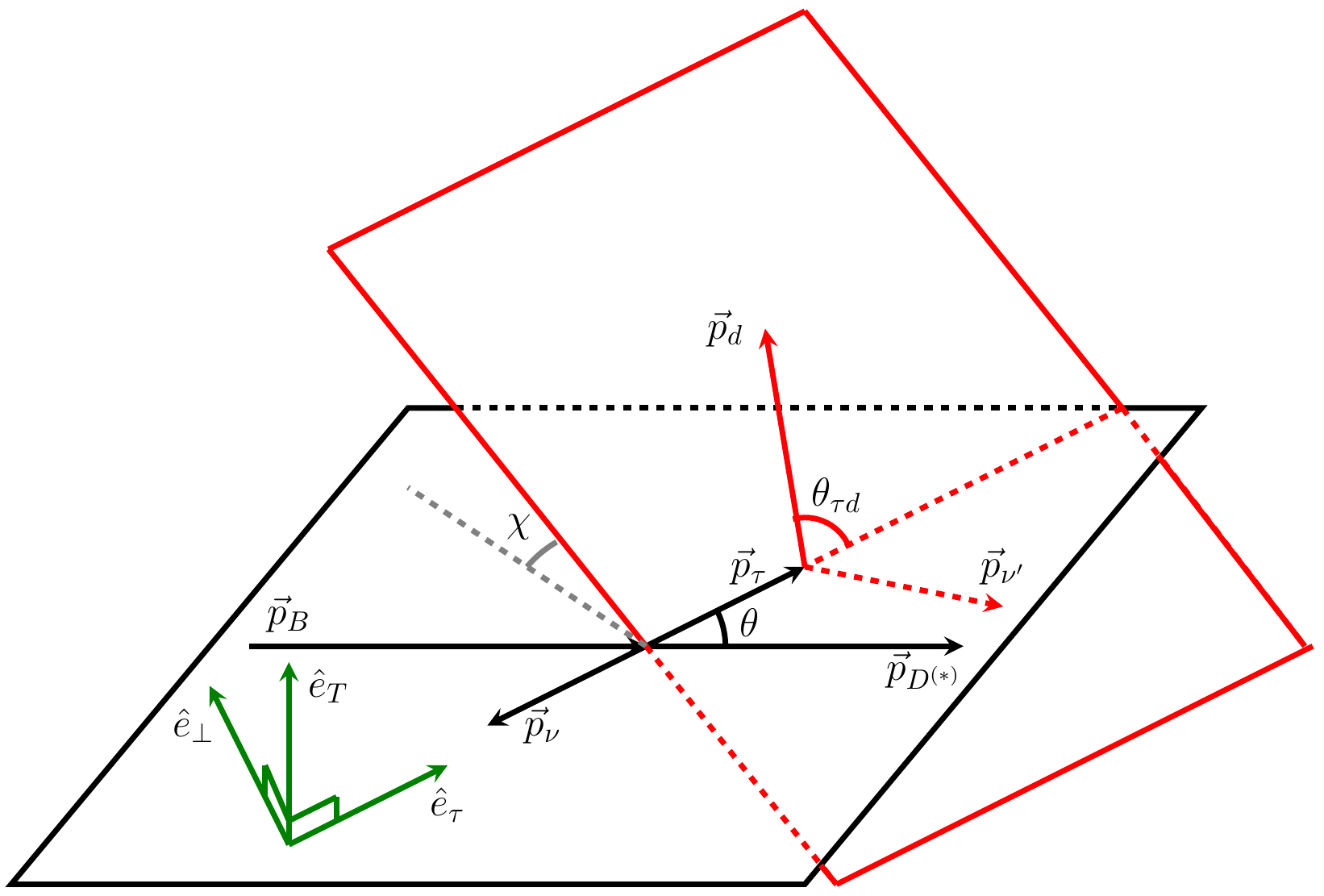}
\caption{The kinematics of $\bar B\to D^{(*)}\tau\nu$  and subsequent $\tau\to d\nu'$ decay processes, in the center-of-mass frame of the leptonic system (the ``$q^2$ frame"). 
The black plane indicates the original decay plane, defined by the $B$ momentum $\vec{p}_B$ (or the $D^{(*)}$ momentum ${\vec p_{D^{(*)}}}$) and the leptonic pair. The red plane is the decay plane of the $\tau$, defined by the visible daughter meson $d$ and invisible daughter neutrino $\nu'$ of the $\tau$. The three directions in which we will project the $\tau$ polarization asymmetries are indicated in green. 
}
\label{fig:kinematics}
\end{figure}

The relevant models for $R_{D^{(*)}}$ and their predictions for these ratios were reviewed in the previous section. However, one can extract more information from the decay processes than just the total decay rate and the ratios $R_{D^{(*)}}$. Shown in Fig.~\ref{fig:kinematics} is a diagram of the detailed kinematics of the decay process. Many of these angles and momenta can be measured or reconstructed, and they provide a much finer probe of the effective Hamiltonian responsible for the decay.  

In particular, using the event kinematics, we can construct asymmetry observables which are sensitive to the different Wilson coefficients in \eqref{Heff}. Four such observables are the forward-backward asymmetry of the $\tau$ lepton with respect to $\vec p_{D^{(*)}}$ in Fig.~\ref{fig:kinematics}, denoted by $\mathcal{A}_{FB}^{(*)}$, and its polarization asymmetry in all three of the $\hat{e}$ directions in Fig.~\ref{fig:kinematics}, denoted by $\mathcal{P}_{\hat{e}}^{(*)}$. 
All of these asymmetries are defined in the  leptonic center of mass frame, which we will also refer to as the ``$q^2$ frame", where $q=p_B-p_{D^{(*)}}=p_\tau+p_\nu$ denotes the four-momentum transferred to the leptonic system by the decaying $B$ meson. As we will see, models with LH and RH neutrinos have a qualitatively different contribution to these asymmetry observables.

We will calculate the dependence of these observables on all the Wilson coefficients in \eqref{Heff} and report the result in the form of numerical formulas (like \eqref{eq:allCRD} for $R_D$ and $R_{D^{*}}$). In particular, we carry out the calculation including the contribution of the operators with right-handed sterile neutrinos with negligible masses compared to the other energy scales in the decay. Full analytic versions are available in the appendices. Wherever possible, we have checked that parts of our calculations (results from the numerical equations, $q^2$ distributions, the SM predictions, etc.) are in agreement with previous studies, e.g. \cite{Bardhan:2016uhr,Ivanov:2017mrj,Tanaka:2012nw,Ivanov:2016qtw}. 
A further consistency check is that the numerical equations for the observables will manifest a symmetry between left- and right-handed neutrinos such that by applying the following transformations,
\bea
\label{eq:transform}
h_\tau \rightarrow - h_\tau,\qquad  C^{S,T}_{LL} \leftrightarrow \left(C^{S,T}_{RR}\right)^*  &,& \qquad C^X_{RL} \leftrightarrow \left(C^X_{LR}\right)^*, \\
 1+ C^{V}_{LL} \leftrightarrow \left(C^{V}_{RR}\right)^*  &,&
\eea
(where $h_\tau$ refers to the $\tau$ helicity) the observables will transform as
\begin{eqnarray}
\label{eq:transform2}
R_{D^{(*)}} \rightarrow R_{D^{(*)}},~~~ \mathcal{P}_x \rightarrow - \mathcal{P}_x ,~~~ \mathcal{A}_{FB} \rightarrow \mathcal{A}_{FB}.
\end{eqnarray}
In writing $1+C^{V}_{LL}$ in \eqref{eq:transform} (and in all the up-coming numerical equations), we are explicitly separating the contribution of the SM operator.\footnote{The complex conjugate in the way the Wilson coefficients are transformed is only relevant for the study of $\mathcal{P}^{(*)}_T$ observables and is essentially an artifact of the definition in \eqref{eq:polzMs} and how the $\tau$ spin transforms under this symmetry. } These symmetries indicate that if we flip the spin of all the external particles and the associated Wilson coefficients, we should get the same result for the decay rate in a particular $q^2$ and $\theta$ direction.  
The interference between the SM term in \eqref{eq:transform} and the sign flip in \eqref{eq:transform2} are the two sources of the qualitatively different contributions from different types of neutrinos.

\subsection{Forward-backward Asymmetry}
\label{subsec:AFB}

The first observable of interest is the forward-backward asymmetry in the $\tau$ lepton decay with respect to the $D^{(*)}$ direction. This observable and its correlation with $R_{D^{(*)}}$ have been studied previously \cite{Bardhan:2016uhr,Duraisamy:2013kcw,Becirevic:2016hea,Alonso:2016gym,Sakaki:2012ft,Datta:2012qk,Ivanov:2015tru,Alok:2016qyh,Ivanov:2017mrj,Alonso:2017ktd}. It is defined as
\begin{equation}
\mathcal{A}_{FB}^{(*)} = \frac{1}{\Gamma^{{(*)}}}  \left(	-	\int_{\theta=0}^{\theta=\pi/2} + \int_{\theta=\pi/2}^{\theta=\pi}							\right) d\theta \frac{d\Gamma^{(*)}}{d\theta},
\label{eq:defAFB}
\end{equation}
where $\theta$ is the angle between the $\tau$  and $D^{(*)}$ momenta in the leptonic system rest frame, see Fig.~\ref{fig:kinematics}, and $\Gamma^{(*)}$ is the total decay rate of $\bar{B} \rightarrow D^{(*)} \tau \nu$. The full analytic expression for $\frac{d\Gamma^{(*)}}{d\theta}$ in terms of all the Wilson coefficients is included in  App.~\ref{sec:analytics}. The numerical formula for $\mathcal{A}_{FB}^{(*)}$ that follows from this is:  
\begin{eqnarray}
\mathcal{A}_{FB}  &\approx & \frac{1}{R_D} \left\lbrace  		-0.11 \left( \left| 1+C^V_{LL} + C^V_{RL}			\right|^2  + \left| C^V_{RR} + C^V_{LR}			\right|^2		\right)				\nonumber \right. \\ 
\label{eq:numericRDAFB}
&-&   0.35 \mathcal{R}e \left[ 	(C^S_{LL}+C^S_{RL}) (C^T_{LL})^*		+   (C^S_{RR}+C^S_{LR})^* (C^T_{RR})		\right] \nonumber \\
&-&    0.24 \mathcal{R}e \left[ 	(1+C^V_{LL}+C^V_{RL}) (C^T_{LL})^*		+   (C^V_{RR}+C^V_{LR})^* (C^T_{RR})		\right]  \nonumber \\
&-&  \left.  0.15 \mathcal{R}e \left[ 	(1+C^V_{LL}+C^V_{RL}) (C^S_{LL}+C^S_{RL})^*		+   (C^V_{RR}+C^V_{LR})^* (C^S_{RR}+C^S_{LR})		\right] \right\rbrace , \nonumber \\
\\
\mathcal{A}_{FB}^* &\approx & \frac{1}{R_{D^*}} \left\lbrace		-0.813 \left(	\left|		C^T_{LL}	\right|^2 + \left|		C^T_{RR}	\right|^2		\right)				\right.		 \nonumber \\
&+&	0.016  \left(	\left|		1+C^V_{LL}			\right|^2	 + \left|		C^V_{RR}			\right|^2		\right) - 0.082 \left(	\left|		C^V_{RL}			\right|^2	 + \left|		C^V_{LR}			\right|^2		\right)  \nonumber \\
&+& 0.066 \mathcal{R}e \left[	C^V_{RL} (1+C^V_{LL})^*  +  (C^V_{LR})^* C^V_{RR}	\right] \nonumber \\
&+& 0.095 \mathcal{R}e \left[	(C^S_{RL} - C^S_{LL}) (C^T_{LL})^*  +  (C^S_{LR} - C^S_{RR})^* C^T_{RR}		\right] \nonumber \\
&+& 0.395 \mathcal{R}e \left[	(1+C^V_{LL} - C^V_{RL}) (C^T_{LL})^*  +  (C^V_{RR} - C^V_{LR})^* (C^T_{RR})	\right]	\nonumber \\
&+& 0.023 \mathcal{R}e \left[	(C^S_{LL} - C^S_{RL} ) (1+C^V_{LL}-C^V_{RL})^*   +    (C^S_{RR} - C^S_{LR} )^* (C^V_{RR}-C^V_{LR}) 		\right] \nonumber \\
&-& \left. 0.142 \mathcal{R}e \left[	(C^T_{LL}) (1+C^V_{LL}+C^V_{RL})	^*  +    (C^T_{RR})^* (C^V_{RR}+C^V_{LR})	\right]	\right\rbrace ,	\nonumber 
\end{eqnarray}
The factor of $R_{D^{(*)}}$ in the denominators are the result of normalizing to the total decay rate $\Gamma^{(*)}$ in \eqref{eq:defAFB}.

\subsection{Tau Polarization Asymmetries}
\label{subsec:PL}

Our second set of observables is comprised of the different polarization asymmetries of the $\tau$ lepton in the decay. Such asymmetries are defined as 
\begin{equation}
\mathcal{P}_{\hat{e}}^{(*)} = \frac{\Gamma_{+\hat{e}}^{(*)} - \Gamma_{-\hat{e}}^{(*)}}{\Gamma_{+\hat{e}}^{(*)} + \Gamma_{-\hat{e}}^{(*)}},
\label{eq:defPt}
\end{equation}
where $\pm$ refer to the two possible outcomes of measuring $\tau$ spin along direction $\hat{e}$. The vector $\hat{e}$ can be in any arbitrary direction. We consider the three directions \cite{Alonso:2017ktd},
\begin{equation}
\hat{e}_\tau = \frac{\vec{p}_\tau}{|\vec{p}_\tau|}, \hspace{0.3 in } \hat{e}_T = \frac{\vec{p}_{D^{(*)}} \times \vec{p}_\tau }{|\vec{p}_{D^{(*)}} \times \vec{p}_\tau|}, \hspace{0.3in} \hat{e}_\perp = \hat{e}_T \times \hat{e}_\tau,
\label{eq:taudirection}
\end{equation}
where $\vec{p}_\tau$ ($\vec{p}_{D^{(*)}}$) is the spatial momentum of the $\tau$ ($D^{(*)}$) in the final state (all in the $q^2$ frame). $\mathcal{P}_{\tau}^{(*)}$ indicates the polarization asymmetry along the longitudinal direction of the $\tau$ lepton, and $\mathcal{P}_{\perp}^{(*)}$ the asymmetry in the decay plane and perpendicular to $\vec{p}_\tau$, while  $\mathcal{P}_{T}^{(*)}$ is the polarization asymmetry along the direction normal to the decay plane including $\tau$ and $D^{(*)}$, see Fig.~\ref{fig:kinematics}. The first two are CP-even while the latter is CP-odd. The details of calculating each $\mathcal{P}_{\hat{e}}^{(*)}$ and their analytic results are included in App.~\ref{sec:analytics}.

\subsubsection{Longitudinal polarization}

The numerical expression for the contribution of all the Wilson coefficients to $\mathcal{P}_{\tau}^{(*)}$ is:
\begin{eqnarray}
\mathcal{P}_{\tau}  &\approx & \frac{1}{R_D} \left\lbrace  		0.402 \left( \left| C^S_{LL} + C^S_{RL}			\right|^2  - \left| C^S_{RR} + C^S_{LR}			\right|^2		\right)		\right.		\nonumber \\ 
\label{eq:numericPtau}
&+&   0.013  \left[ \left|C^T_{LL}\right|^2		-  \left|C^T_{RR}\right|^2	\right]  + 0.097  \left[ \left| 1+C^V_{LL}+C^V_{RL}\right|^2		-  \left| C^V_{RR}+C^V_{LR}\right|^2	\right] 		 \nonumber \\
&+&    0.512 \mathcal{R}e \left[ 	(1+C^V_{LL}+C^V_{RL}) (C^S_{LL}+C^S_{RL})^*		-		(C^V_{RR}+C^V_{LR})^* (C^S_{RR}+C^S_{LR})	\right]  \nonumber \\
&-&   \left. 0.099 \mathcal{R}e \left[ 	(1+C^V_{LL}+C^V_{RL}) (C^T_{LL})^*		-   (C^V_{RR}+C^V_{LR})^* (C^T_{RR})		\right] \right\rbrace  \nonumber \\
\\
\mathcal{P}_{\tau}^* &\approx & \frac{1}{R_{D^*}} \left\lbrace		-0.127 \left(	\left|		1+ C^V_{LL}	\right|^2 + \left|		C^V_{RL}	\right|^2 - \left|		C^V_{RR}	\right|^2 - \left|		C^V_{LR}	\right|^2 		\right)				\right.		 \nonumber \\
&+&	0.011  \left(	\left|		C^S_{LL}	- C^S_{RL}			\right|^2	 - \left| C^S_{RR}	- C^S_{LR}	\right|^2			\right)   + 0.172 \left(		\left|	 C^T_{LL}   \right|^2		- \left|	 C^T_{RR}   \right|^2	\right)\nonumber \\
&+& 0.031 \mathcal{R}e\left[	\left(1+C^V_{LL}-C^V_{RL}	\right)	\left(C^S_{RL}-C^S_{LL}	\right)^*	-   \left(C^V_{RR}-C^V_{LR}	\right)^*	\left(C^S_{LR}-C^S_{RR}	\right)		\right] \nonumber \\
&+& 0.350 \mathcal{R}e \left[		\left(	1+C^V_{LL}		\right)	(C^T_{LL})^*  -   \left(	C^V_{RR}		\right)^*	(C^T_{RR})  \right]  \nonumber \\
&-&\left. 0.481 \mathcal{R}e\left[		(C^V_{RL}) (C^T_{LL})^*  -  ( C^V_{LR})^* (C^T_{RR})	\right]	+ 0.216  \mathcal{R}e \left[	(1+C^V_{LL})(C^V_{RL})^*   -  (C^V_{RR})^*(C^V_{LR})	\right]\right\rbrace . \nonumber 
\end{eqnarray}

\subsubsection{Perpendicular polarization}
\label{subsec:Pperp}

Similar to the previous section we include the numerical expression for contribution of all the Wilson coefficients to $\mathcal{P}_{\perp}^{(*)}$.
\begin{eqnarray}
\mathcal{P}_{\perp}  &\approx & \frac{1}{R_D} \mathcal{R}e \left\lbrace  		- 0.350 \left[ (C^T_{LL}) \left( C^S_{LL} + C^S_{RL}			\right)^*  - (C^T_{RR})^* \left( C^S_{RR} + C^S_{LR}			\right)		\right]			\right.	\nonumber \\ 
&-&   0.357  \left[ \left(	1+C^V_{LL} + C^V_{RL}	\right)  \left(	C^S_{LL} + C^S_{RL}	\right)^*	-   \left(	C^V_{RR} + C^V_{LR}	\right)^*	\left(	C^S_{RR} + C^S_{LR}	\right) \right] 		 \nonumber \\
&-&    0.247  \left[ 	(1+C^V_{LL}+C^V_{RL})^* (C^T_{LL}) 	-		(C^V_{RR}+C^V_{LR}) (C^T_{RR})	^*  \right]  \nonumber \\
&-& \left.   0.250  \left[ 	\left| 1+C^V_{LL}+C^V_{RL}\right|^2		-   	\left| C^V_{RR}+C^V_{LR}\right|^2			\right] \right\rbrace  \nonumber \\
\label{eq:numericPperp}
\\
\mathcal{P}_{\perp}^* &\approx & \frac{1}{R_{D^*}} \mathcal{R}e \left\lbrace		\left(	C^S_{RR}-C^S_{LR}	\right)  \left[ 0.099 C^T_{RR} - 0.054 \left(	C^V_{RR} - C^V_{LR}	\right) \right]^*	 \right.  \nonumber \\
&-& \left(	C^S_{LL}-C^S_{RL}	\right)^*  \left[ 0.099 C^T_{LL} - 0.054 \left(1+	C^V_{LL} - C^V_{RL}	\right) \right] \nonumber \\
&+& (C^T_{RR}) \left[	0.146 	C^V_{RR} - 0.478 C^V_{LR}  -  1.855 C^T_{RR} 	\right]^*	 \nonumber \\
&-&  (C^T_{LL})^* \left[	0.146 	(1+C^V_{LL}) - 0.478 C^V_{RL}  -  1.855 C^T_{LL} 	\right]	 \nonumber \\
&+&  (C^V_{LR}) \left[	-0.081 C^T_{RR} + 0.025 C^V_{LR} -0.075 C^V_{RR}		\right]^*  \nonumber \\
&-&  (C^V_{RL})^* \left[	-0.081 C^T_{LL} + 0.025 C^V_{RL} -0.075 (1+C^V_{LL}	)	\right]  \nonumber \\
&+&  (C^V_{RR}) \left[	-0.071 C^T_{RR} -0.075 C^V_{LR} +0.126 C^V_{RR}		\right]^*  \nonumber \\
&-&  \left. (1+C^V_{LL})^* \left[	-0.071 C^T_{LL} -0.075 C^V_{RL}+ 0.126 (1+C^V_{LL})		\right] \right\rbrace . \nonumber 
\end{eqnarray}

\subsubsection{Transverse polarization}
\label{subsec:PT}

Finally, we present the numerical formulas for $\mathcal{P}_{T}^{(*)}$:
\begin{eqnarray}
\mathcal{P}_{T}  &\approx & \frac{1}{R_D} \mathcal{I}m \left\lbrace  		- 0.350 \left[ (C^T_{LL}) \left( C^S_{LL} + C^S_{RL}			\right)^*  - (C^T_{RR})^* \left( C^S_{RR} + C^S_{LR}			\right)		\right]			\right.	\nonumber \\ 
&-&   0.357  \left[ \left(	1+C^V_{LL} + C^V_{RL}	\right)  \left(	C^S_{LL} + C^S_{RL}	\right)^*	-   \left(	C^V_{RR} + C^V_{LR}	\right)^*	\left(	C^S_{RR} + C^S_{LR}	\right) \right] 		 \nonumber \\
&-&  \left.  0.247  \left[ 	(1+C^V_{LL}+C^V_{RL})^* (C^T_{LL})  	-		(C^V_{RR}+C^V_{LR}) (C^T_{RR})	^*  \right]  \right\rbrace  \nonumber \\
\label{eq:numericPT}
\\
\mathcal{P}_{T}^* &\approx & \frac{1}{R_{D^*}} \mathcal{I}m \left\lbrace		\left(	C^S_{RR}-C^S_{LR}	\right)  \left[ 0.099 C^T_{RR} - 0.054 \left(	C^V_{RR} - C^V_{LR}	\right) \right]^*	 \right.  \nonumber \\
&-& \left(	C^S_{LL}-C^S_{RL}	\right)^*  \left[ 0.099 C^T_{LL} - 0.054 \left(1+	C^V_{LL} - C^V_{RL}	\right) \right] \nonumber \\
&+& (C^T_{RR}) \left[	0.146 	C^V_{RR} - 0.478 C^V_{LR} 	\right]^*	 -  (C^T_{LL})^* \left[	0.146 (1+	C^V_{LL} ) - 0.478 C^V_{RL}  \right]	  \nonumber \\
&-&  (C^V_{LR}) \left[	0.081 C^T_{RR} 	\right]^*  + (C^V_{RL})^* \left[	0.081 C^T_{LL} 	\right]   \nonumber \\
&-&  \left. (C^V_{RR}) \left[	0.071 C^T_{RR} 	\right]^* + (1+C^V_{LL})^* \left[	0.071 C^T_{LL}  	\right] \right\rbrace  \nonumber 
\end{eqnarray}

The $\mathcal{P}_{T}^{(*)}$ observables are particularly interesting to measure as they can provide us with a way to hunt for CP-violation in $B$-meson decays. 
The SM prediction for these observables is zero. In this work we focus on the $\mathcal{P}_{T}^{(*)}$ observables for the $\bar{B}$ meson decay. Due to its CP-odd nature, the associated observables in the decay of $B$ mesons can be obtained by complex conjugation of all the Wilson coefficients, i.e. an overall sign.

\subsection{Overview of the Experimental Results and Proposals}
\label{subsec:exp}

\begin{figure}
\includegraphics[scale=0.6]{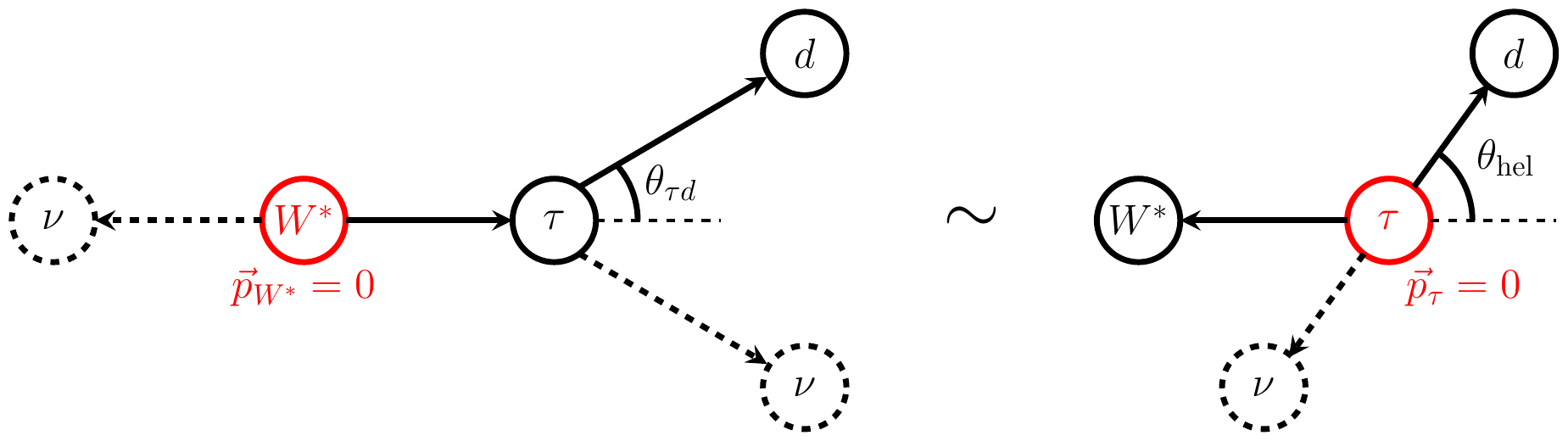}
\caption{ A schematic showing the Lorentz boost that relates the angles $\theta_{\tau d}$ in the $q^2$ frame on the left and $\theta_{\mathrm{hel}}$ in the $\tau$ rest frame on the right. The former angle is reconstructible at the $B$-factories, while the latter is used to extract $\mathcal{P}_\tau^{(*)}$. 
Although the $\tau$ momentum vector cannot be fully reconstructed at the $B$ factories, its magnitude is measurable, and this is sufficient to relate the two frames.
}
\label{fig:tau}
\end{figure}

So far the only asymmetry observable studied experimentally is ${\mathcal P}_\tau^*$, by Belle in a series of works \cite{Abdesselam:2016xqt,Hirose:2016wfn,PhysRevD.97.012004}. The missing energy in these decays prevents us from fully reconstructing all the momenta and thus complicates the measurement of different angular observables. However, Belle was able to extract ${\mathcal P}_\tau^*$ from single-prong $\tau$ decays, $\tau \rightarrow d \nu$ with $d= \pi , \rho$, using the observation that the differential decay rate of $\bar{B}\rightarrow D^{*} \tau \nu$, $\tau\to d\nu$ can be written as 
\begin{equation} 
\frac{1}{\Gamma} \frac{d\Gamma}{d\theta_{\rm hel}} =  \frac{1}{2}\left(		1 + \alpha_d \mathcal{P}_\tau^{*} \cos \theta_{\rm hel}		\right),
\label{eq:thetahelPtau}
\end{equation}
where $\theta_{\rm hel}$ is the angle between $d$ and the opposite of the $W^*$ direction in the $\tau$ rest frame, see Fig.~\ref{fig:tau}. The constant $\alpha_d$ captures the sensitivity to $\mathcal{P}_\tau^{*}$ of the particular $\tau$ decay channel under study.

Unfortunately, the $\tau$ rest frame is not reconstructible, even at the $B$-factories. What is reconstructible is the $q^2$ frame, i.e.~the leptonic center of mass frame, by boosting to the frame where the (fully measurable) $B$ and $D^*$ momenta are pointed in the same direction. Furthermore, in the $q^2$ frame, the angle $\theta_{\tau d}$ between the $\tau$ and its daughter meson $d$ is given by 
\begin{equation}
\cos \theta_{\tau d} = \frac{2E_\tau E_d-m_\tau^2-m_d^2}{2|\vec{p}_\tau||\vec{p}_d|}.
\label{eq:thetataud}
\end{equation}
The RHS is completely known, because the magnitude of the $\tau$ momentum is a function of $q^2$ in the $q^2$ frame  
\beq
 |\vec{p}_\tau| = {q^2-m_\tau^2\over 2\sqrt{q^2}}.
\eeq

As evident from Fig.~\ref{fig:tau}, the angle $\theta_{\tau d}$ is related to $\theta_{\mathrm{hel}}$ via a boost along the $\tau$ momentum direction. Although we do not know the direction, it is enough to know the magnitude:
\begin{equation}
|\vec{p^\tau_d} |\cos \theta_{\rm hel} = - \gamma \frac{|\vec{p}_\tau |}{E_\tau} E_d + \gamma |\vec{p}_d| \cos \theta_{\tau d},
\label{eq:thetahel}
\end{equation}
where $|\vec{p^\tau_d} | = (m_\tau^2-m_d^2)/(2m_\tau)$ is the momentum of the daughter meson in the $\tau$ rest frame, and $\gamma = E_\tau/m_\tau$. This relation determines $\theta_{\mathrm{hel}}$ in terms of all measurable quantities, and allowed Belle to obtain a measurement of ${\mathcal P}_\tau^*=-0.38\pm0.51^{+0.21}_{-0.16}$ (compared to a SM prediction of $({\mathcal P}_\tau^*)_{{\rm SM}}=-0.497$).

Although this method works, it resulted in an enormous uncertainty, and has so far only been applied to  ${\mathcal P}_\tau^*$. 
There are further proposals in the literature on how we can infer additional asymmetry observables from the angular distribution of the visible daughter mesons in the $\tau$ lepton decays. In particular, \cite{Alonso:2017ktd} puts forward methods for measuring $\mathcal{P}_{\tau}$, $\mathcal{P}_{\perp}$, and $\mathcal{A}_{FB}$ in $B\to D\tau\nu$ decays (with $\tau\to d\nu)$, claiming a better attainable precision than the Belle procedure described above. 

In their method, $q^2$, $E_d$, and the angle $\theta_d$ between $d$ and $D$ -- all evaluated in the $q^2$ frame, and all directly measurable -- are used to express the differential decay rates, 
\begin{equation}
\frac{d^3\Gamma}{dq^2 dE_d d\cos \theta_d} = \mathcal{B}_d \frac{\mathcal{N}}{2m_\tau} \left( 	I_0 \left(q^2,E_d\right) + I_1 \left(q^2,E_d\right) \cos \theta_d + I_2 \left(q^2,E_d\right) \cos^2 \theta_d	\right),
\label{eq:alonsometh1}
\end{equation}
where $\mathcal{B}_d$ is the branching ratio of $\tau$ into the daughter meson under study, $\mathcal{N}$ is a normalization factor, and $I_{0,1,2}$ are functions of $q^2$ and $E_d$ defined in \cite{Alonso:2017ktd}. After integrating over $\theta_d$, adding together or subtracting the decay rates into the two spatial hemispheres give rise to double distributions, from which $\mathcal{P}_\tau$, $\mathcal{P}_\perp$ and $\mathcal{A}_{FB}$ can be extracted. 

In Tab.~\ref{tab:observables}, we list the projected Belle~II sensitivity claimed in \cite{Alonso:2017ktd} (which we also adopt in this work), as well as our calculation for the SM predictions.  Although there are currently no analogous proposals to measure the $D^*$ asymmetry observables in the literature, we believe that a similar method to the one proposed in \cite{Alonso:2017ktd} should be applicable. 

At present, there is no substantiative experimental proposal for how to measure $\mathcal{P}_{T}^{(*)}$ at Belle II.\footnote{In \cite{Ivanov:2017mrj} it has been shown that the total decay rates above and their dependence on the azimuthal angle $\chi$ between the two planes in Fig.~\ref{fig:kinematics} contains information about $\mathcal{P}_T^{(*)}$. We cannot confirm the claim that this angle is experimentally accessible and are not aware of any experimental proposals for its measurement at Belle~II.} However, we have included it in our study, owing to the important role it can play in distinguishing certain models from one another (see the next section), and in the hopes that viable proposals for how to measure it will emerge in the future. 

\begin{table}
\centering
\resizebox{\columnwidth}{!}{
\begin{tabular}{|c|c|c|c|c|c|c|c|c|}
\hline 
Observable & $\mathcal{A}_{FB}$ & $\mathcal{A}^*_{FB}$ & $\mathcal{P}_{\tau}$ & $\mathcal{P}_{\tau}^*$ & $\mathcal{P}_{\perp}$ & $\mathcal{P}_{\perp}^*$ & $\mathcal{P}_{T}$ & $\mathcal{P}_{T}^*$ \\ 
\hline 
SM value & $-0.360$ & 0.063 & 0.325 & $-0.497$  & $-0.842$ & $-0.499$ & 0 & 0 \\ 
\hline 
Projected Precision \cite{Alonso:2017ktd} & $10\%$  & $-$ & $3\%$  & $-$ & $10\%$  & $-$ & $-$ & $-$ \\ 
\hline 
\end{tabular} 
}
\caption{Observables studied in this work, our numerical calculation for the prediction in the SM, and the projected Belle~II sensitivity (assuming the 50~ab$^{-1}$ full data set) where available. We use these observables to identify different explanations of the anomalies. In the upcoming sections we will assume the observables in $B \rightarrow D^* \tau \nu$ are measured with the same uncertainty as in $B \rightarrow D \tau \nu$.}
\label{tab:observables}
\end{table}

\section{Distinguishing Different Solutions}
\label{sec:discerning}

Having calculated these asymmetry observables, we now use them to distinguish between different simplified models for the $R_{D^{(*)}}$ anomalies (see Sec.~\ref{sec:models}). As the range of possible Wilson coefficients depends on the value of $R_D$ and $R_{D^*}$ after the Belle~II data set is collected, we consider the two benchmark scenarios described in Sec.~\ref{subsec:scenarios} and indicated in Figs.~\ref{fig:rdrdsrange_operators} and \ref{fig:rdrdsrange_models}.

\subsection{$10\sigma$ Scenario}
\label{subsec:comparison1}

In this scenario, for the models involving the LH neutrinos, the LQ $U_1$, as well as the single operators $\mathcal{O}^T_{LL}$ and $\mathcal{O}^V_{RL}$, will be able to explain the anomalies while satisfying the experimental bounds mentioned above. Among the RH neutrino proposals, only $U_1$ LQ will remain viable. 

\begin{figure}
\includegraphics[width=\columnwidth]{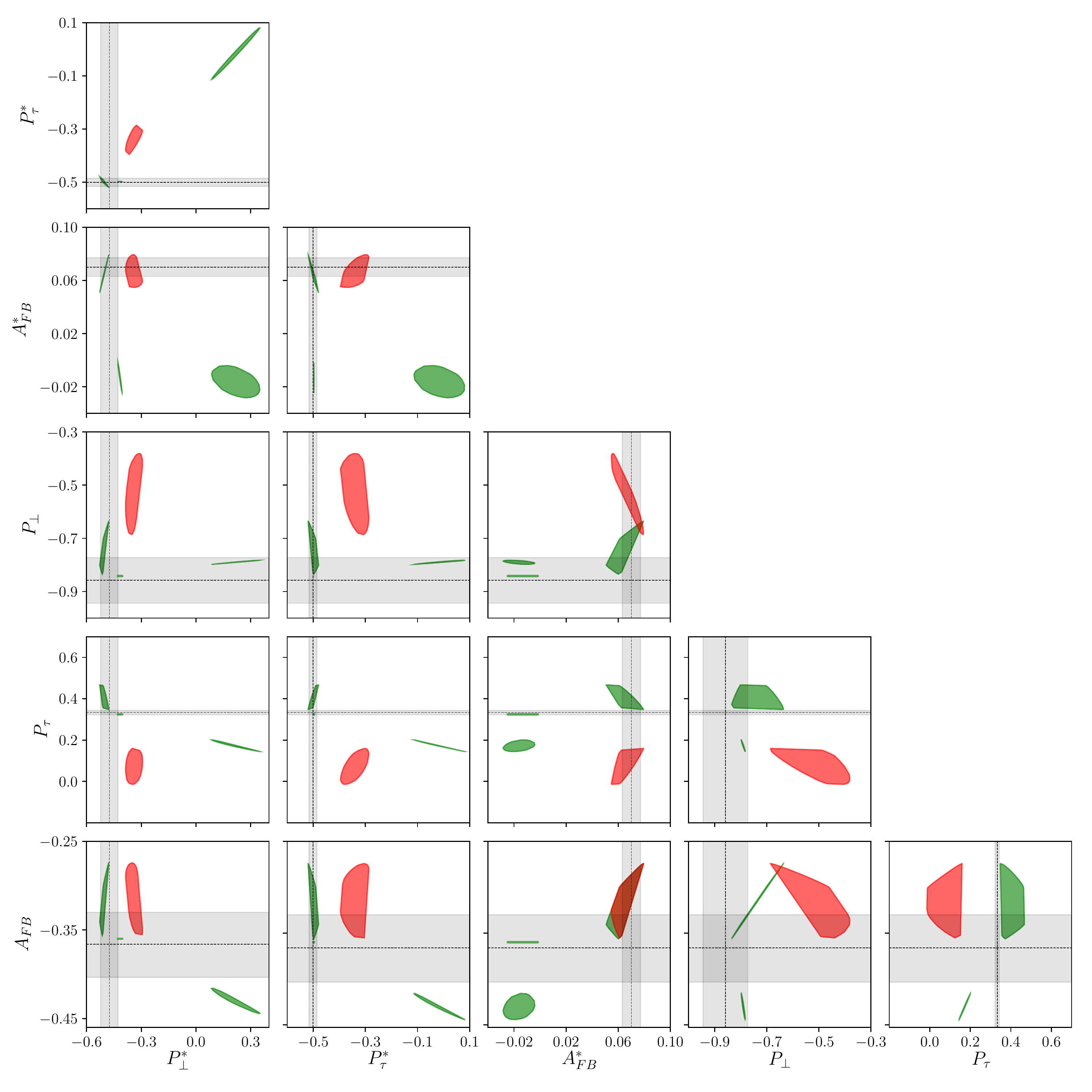}
\caption{Two-dimensional plots of asymmetry observables for the $10\sigma$ scenario. We scan over Wilson coefficients that result in $R_{D^{(*)}}$ values within the $2\sigma$ Belle~II error ellipse centered on the present-day world averages.
We also impose the $Br(B_c \rightarrow \tau \nu)\leqslant 10 \%$ bound \cite{Akeroyd:2017mhr}. The projected Belle~II precision for each observable, centered on the SM prediction, is indicated by the dashed gray lines, see the text. Regions which can be realized by models with LH SM neutrinos (shown in green) are from $U_1$ LQ and single operators $\mathcal{O}^T_{LL}$ and $\mathcal{O}^V_{RL}$, while the one requiring new RH neutrinos (shown in red) corresponds to $U_1$ LQ. We can distinguish all the models from one another by measuring these asymmetry observables. }
\label{fig:scan_optimistic}
\end{figure}

Fig.~\ref{fig:scan_optimistic} shows the ranges of CP-even asymmetry observables that are achievable in each model, projected here into 2D plots, one for each pair of observables. In each model, we have scanned over the (complex) Wilson coefficients of the model, subject to the following constraints: $R_D$ and $R_{D^*}$ should be within the 2$\sigma$ Belle~II error ellipse for this scenario;  $Br(B_c \rightarrow \tau \nu) < 10\%$. The gray regions in each plot denote the Belle~II projected relative uncertainty from Tab.~\ref{tab:observables} centered around the SM prediction; for the observables in the $\bar{B}\rightarrow D^* \tau \nu$ process, as there are no available projection, we assume the same relative uncertainties as in the $\bar{B}\rightarrow D \tau \nu$ decay.
Models affected by the $b\rightarrow s \nu\nu$ bounds are already ruled out in this scenario and are not included in Fig.~\ref{fig:scan_optimistic}. Further details on how to efficiently carry out this scan are included in App.~\ref{sec:scan}. 

It is obvious from Fig.~\ref{fig:scan_optimistic} that by measuring all these observables we can distinguish well each individual model. In particular, the observables $\mathcal{P}^{(*)}_\tau$ and $\mathcal{A}_{FB}^*$ are the most promising discriminators. This conclusion would remain unchanged even if we had applied the looser $Br(B_c \rightarrow \tau \nu) < 30\%$ bound.

\subsection{$5\sigma$ Scenario}
\label{subsec:comparison2}

In our second scenario for the outcome of Belle II measurements, we study the situation in which the observed values of the $R_{D^{(*)}}$ anomalies in the Belle~II data are reduced significantly from the present average, but still significant enough to be claimed as a $5\sigma$ discovery, see Section \ref{subsec:scenarios} for details. With the reduced values of $R_{D^{(*)}}$, many more models become viable. The minimal models with the $R_2$, $S_1$, or $U_1$ LQs, as well as the individual operators $C^T_{LL}$ and $C^V_{RL}$ can explain the anomalies with a LH neutrino in the decay. For the solutions with the RH neutrinos, a $U_1$ or a $S_1$ LQ, a $W'$, or the single operator $C^V_{LR}$ are viable. Given the severe constraints from the $b\rightarrow s\nu\nu$ processes, the $S_1$ LQ and the $W'$ mediators coupled to RH neutrinos are degenerate; since these two mediators generate a subset of operators generated by a $U_1$ LQ ($C^V_{RR}$), it is impossible to distinguish these three mediators from one another. 

\begin{figure}
\includegraphics[width=\columnwidth]{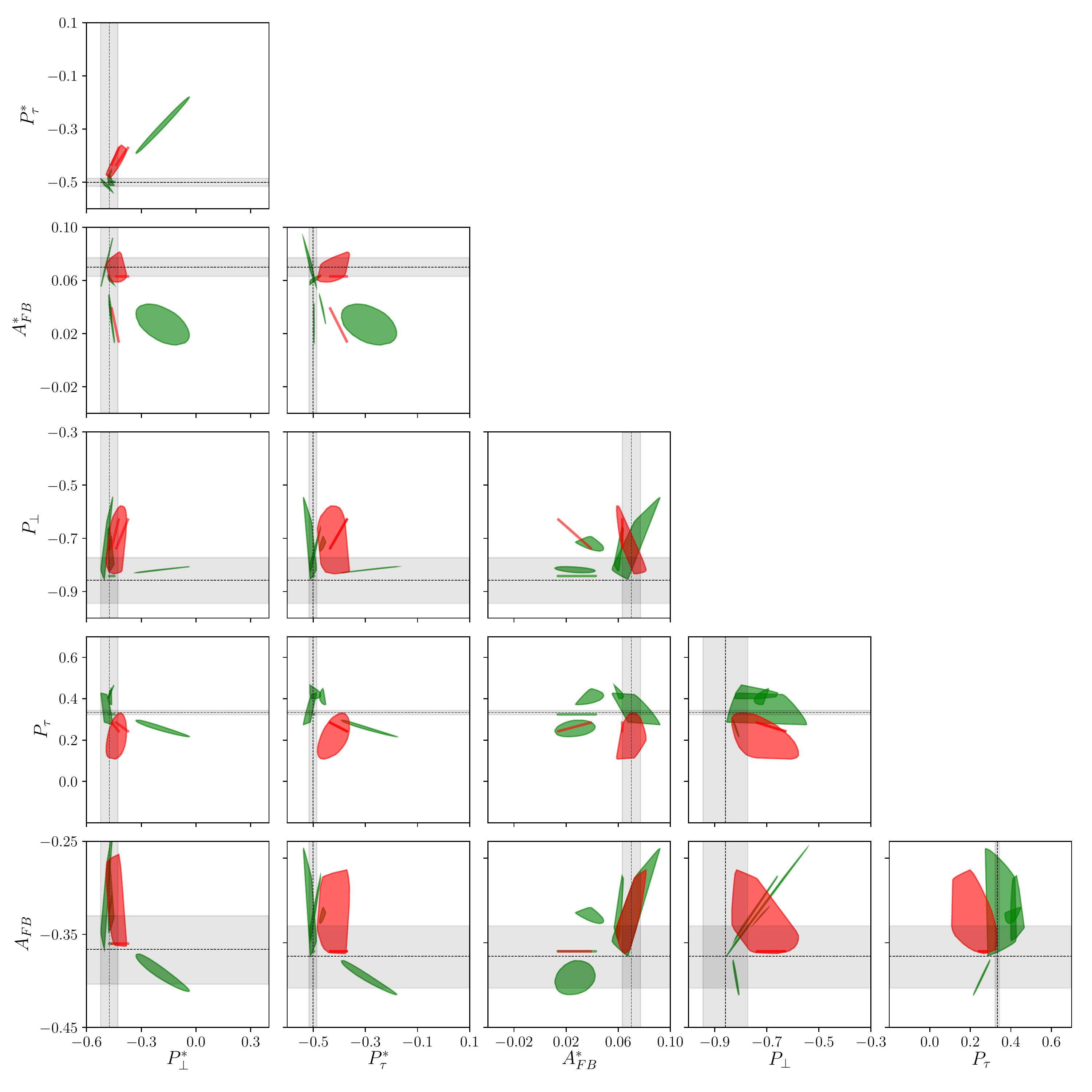}
\caption{Two-dimensional plots of asymmetry observables in the $5\sigma$ scenario. We scan over Wilson coefficients which result in $R_{D^{(*)}}$ values within the $2\sigma$ Belle~II error ellipse centered at $R_D=0.34$ and $R_{D^*}=0.275$.
We also impose the $b\rightarrow s \nu \nu$ bound \cite{Altmannshofer:2009ma,Grygier:2017tzo} and the $Br(B_c \rightarrow \tau \nu)\leqslant 10 \%$ bound \cite{Akeroyd:2017mhr}. The projected Belle~II precision for each observable, centered on the SM prediction, is indicated by the dashed gray lines, see the text. All the currently viable models and single operators remain viable in this scenario. Regions which can be realized by models with LH SM neutrinos are shown in green, while those requiring new RH neutrinos are in red.  }
\label{fig:scan_worst-case}
\end{figure}

Similar 2D plots for this scenario as in the previous one are included in Fig.~\ref{fig:scan_worst-case}. We see immediately that the various regions are much closer together than in Fig.~\ref{fig:scan_optimistic}, as expected from the reduced requirement from $R_{D^{(*)}}$. Combining the results in different plots shows that we can still distinguish different neutrino chiralities (the green regions vs. the red ones) from each other. The best plots that can collectively illustrate this point (again, highly contingent on the projected precision of the Belle~II measurement) are $\mathcal{P}_\tau$--$\mathcal{A}_{FB}^*$, $\mathcal{P}_\tau$--$\mathcal{P}_\tau^*$, $\mathcal{A}_{FB}^*$--$\mathcal{P}_\tau^*$, and $\mathcal{A}_{FB}^*$--$\mathcal{P}_\perp^*$.

While we can still discern models with different neutrino chiralities, it is not immediately obvious if we can distinguish models with the same type of neutrinos from each other. To quantify how well we can separate these models, we use a crude $\chi^2$ measure that includes the six CP-even asymmetry observables from Fig.~\ref{fig:scan_worst-case} as well as the $R_{D^{(*)}}$ ratios themselves, for a total of 8 {\it d.o.f.}.  As there are no data available at this point, the correlation between different asymmetry observables (and $R_{D^{(*)}}$) is not known; we neglect these correlations in this $\chi^2$ estimation and only use the current $\rho_{\rm corr}=-0.2$ between the $R_{D^{(*)}}$ ratios. We calculate this $\chi^2$ between all pairs of scanned points from the models that we want to distinguish.   We use the relative uncertainties from Tab.~\ref{tab:observables} in this calculation; as there are no projections for the precision in measuring the quantities in $\bar{B}\rightarrow D^* \tau \nu$, we use the same relative uncertainty as their counterparts in $\bar{B}\rightarrow D \tau \nu$ from Tab.~\ref{tab:observables} in our calculation.

Our $\chi^2$ estimation corroborates the conclusion from Fig.~\ref{fig:scan_worst-case} that with measuring all CP-even observables we can distinguish models with different neutrinos from each other.\footnote{The closest points from two solutions with different neutrino chiralities belong to the $C^T_{LL}$ and the $C^V_{LR}$ individual operator solutions, with a minimum separation of $2.1$ $\chi^2/d.o.f$. For the heavy mediator solutions the closest pair of points belong to the $U_1$ LQ models coupled to different types of neutrinos, with the minimum separation of $3.4$ $\chi^2/d.o.f$.} However, depending on the outcome of the measurements, it may not be possible to discern individual models with similar neutrino chiralities from one another.

As noted before, the three possible mediators coupled to RH neutrinos ($U_1$ and $S_1$ LQs, and a $W'$) give rise to identical or overlapping parameter spaces, and thus cannot be distinguished from one another on the basis of this effective Hamiltonian alone. Our $\chi^2$ measure, however, indicates that we can distinguish them from the single operator $C^V_{LR}$.  

Our $\chi^2$ estimate further shows that we can distinguish all the viable mediators with LH neutrinos ($U_1$, $S_1$, and $R_2$ LQs) from the individual operators with the same type of neutrinos ($C^T_{LL}$ and $C^V_{RL}$). However, there exist some outcomes where different viable heavy mediators ($U_1$, $S_1$, and $R_2$ LQs) can not be told apart.

\begin{table}
\centering
\resizebox{\columnwidth}{!}{
\begin{tabular}{|c|c|c|c|c|c|c|}
\hline 
\multirow{2}{*}{\small{Model}} & \multirow{2}{*}{$\left(C^S,C^V\right)$ } & \small{$R_D \left[ \pm 0.010 \right]$}  & \small{$\mathcal{A}_{FB}\left[ \pm 0.037\right]$} & \small{$\mathcal{P}_\tau \left[ \pm 0.010 \right]$} & \small{$\mathcal{P}_\perp \left[ \pm 0.086\right]$} & \multirow{2}{*}{\small{$\Delta \chi^2/d.o.f$}}  \\
& &  \small{$R_{D^*} \left[ \pm 0.005\right]$}  & \small{$\mathcal{A}_{FB}^* \left[ \pm 0.007\right]$} & \small{$\mathcal{P}_\tau^* \left[\pm 0.015\right]$} & \small{$\mathcal{P}_\perp^* \left[ \pm 0.048\right]$} & \\ 
\hline 
\hline 
\multirow{2}{*}{$S^L_1$ LQ} & \multirow{2}{*}{$\left(0.062+0.065i,0.005 \right)$ } & 0.333 & $-0.347$ & $0.402$ & $-0.823$ & \multirow{4}{*}{\small{1.14}} \\
& &  0.262 & 0.060 & $-0.510$ & $-0.478$ & \\
\cline{1-6} 
\multirow{2}{*}{$U^L_1$ LQ} & \multirow{2}{*}{$\left(0.041+0.076i,0.017 \right)$ } & 0.332 & $-0.354$ & 0.376 & $-0.830$ & \\
& &  0.263 & 0.060 & $-0.498$ & $-0.514$ & \\
\hline
\end{tabular} 
}
\vskip 0.3cm
\resizebox{\columnwidth}{!}{
\begin{tabular}{|c|c|c|c|c|c|c|}
\hline 
\multirow{2}{*}{\small{Model}} & \multirow{2}{*}{$\left(C^S,C^V\right)$ } & \small{$R_D \left[ \pm 0.010 \right]$}  & \small{$\mathcal{A}_{FB}\left[ \pm 0.037\right]$} & \small{$\mathcal{P}_\tau \left[ \pm 0.010 \right]$} & \small{$\mathcal{P}_\perp \left[ \pm 0.086\right]$} & \multirow{2}{*}{\small{$\Delta \chi^2/d.o.f$}}  \\
& &  \small{$R_{D^*} \left[ \pm 0.005\right]$}  & \small{$\mathcal{A}_{FB}^* \left[ \pm 0.007\right]$} & \small{$\mathcal{P}_\tau^* \left[\pm 0.015\right]$} & \small{$\mathcal{P}_\perp^* \left[ \pm 0.048\right]$} & \\ 
\hline 
\hline 
\multirow{2}{*}{$S^L_1$ LQ} & \multirow{2}{*}{$\left(0.011+0.371i,0.006 \right)$ } & 0.362 & $-0.288$ & $0.441$ & $-0.697$ & \multirow{4}{*}{\small{1.16}} \\
& &  0.265 & 0.061 & $-0.484$ & $-0.466$ & \\
\cline{1-6} 
\multirow{2}{*}{$R_2$ LQ} & \multirow{2}{*}{$\left(-0.002-0.37i,. \right)$ } & 0.352 & $-0.320$ & $0.428$ & $-0.727$ & \\
& &  0.261 & 0.048 & $-0.474$ & $-0.475$ & \\
\hline
\end{tabular} 
}
\vskip 0.3cm
\resizebox{\columnwidth}{!}{
\begin{tabular}{|c|c|c|c|c|c|c|}
\hline 
\multirow{2}{*}{\small{Model}} & \multirow{2}{*}{$\left(C^S,C^V\right)$ } & \small{$R_D \left[ \pm 0.010 \right]$}  & \small{$\mathcal{A}_{FB}\left[ \pm 0.037\right]$} & \small{$\mathcal{P}_\tau \left[ \pm 0.010 \right]$} & \small{$\mathcal{P}_\perp \left[ \pm 0.086\right]$} & \multirow{2}{*}{\small{$\Delta \chi^2/d.o.f$}}  \\
& &  \small{$R_{D^*} \left[ \pm 0.005\right]$}  & \small{$\mathcal{A}_{FB}^* \left[ \pm 0.007\right]$} & \small{$\mathcal{P}_\tau^* \left[\pm 0.015\right]$} & \small{$\mathcal{P}_\perp^* \left[ \pm 0.048\right]$} & \\ 
\hline 
\hline 
\multirow{2}{*}{$U^L_1$ LQ} & \multirow{2}{*}{$\left(0.052+0.113i,0.013 \right)$ } & 0.338 & $-0.350$ & $0.393$ & $-0.819$ & \multirow{4}{*}{\small{1.13}} \\
& &  0.261 & 0.059 & $-0.495$ & $-0.515$ & \\
\cline{1-6} 
\multirow{2}{*}{$R_2$ LQ} & \multirow{2}{*}{$\left(-0.022-0.335i,. \right)$ } & 0.331 & $-0.326$ & 0.396 & $-0.746$ & \\
& &  0.263 & 0.046 & $-0.473$ & $-0.477$ & \\
\hline
\end{tabular} 
}
\caption{
Pairs of benchmark points for the LQ models $S_1$, $U_1$, and $R_2$ coupled to LH neutrinos that are less than $1\sigma$ apart in our estimation. The approximate uncertainties using Tab.~\ref{tab:observables} are quoted in the first row as well. We need further measurements to distinguish these models in these cases. 
}
\label{tab:scenario5}
\end{table}

In Tab.~\ref{tab:scenario5} we list benchmark pairs of measurement outcomes in which pairs of mediators $S_1$, $U_1$, or $R_2$ coupled to LH neutrinos, are not distinguishable. Here we take one of the operators to have real Wilson coefficients without loss of generality, since the observables are insensitive to an overall rephasing of all the Wilson coefficients. The $C^S$ and $C^V$ refer to different scalar and vector Wilson coefficients for the different models; see Tab.~\ref{tab:operators} for details. 

\begin{figure}
\resizebox{\columnwidth}{!}{
\includegraphics[scale=0.3]{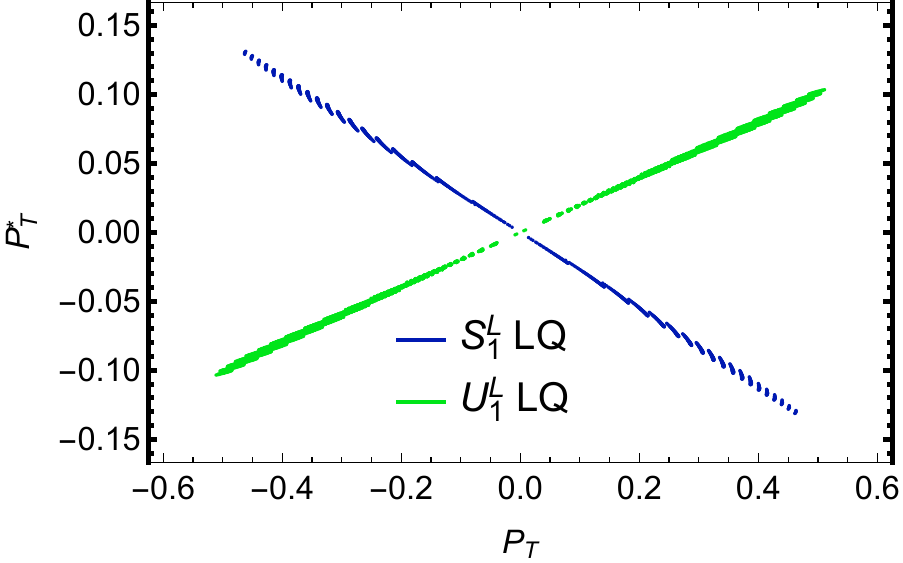}
\includegraphics[scale=0.3]{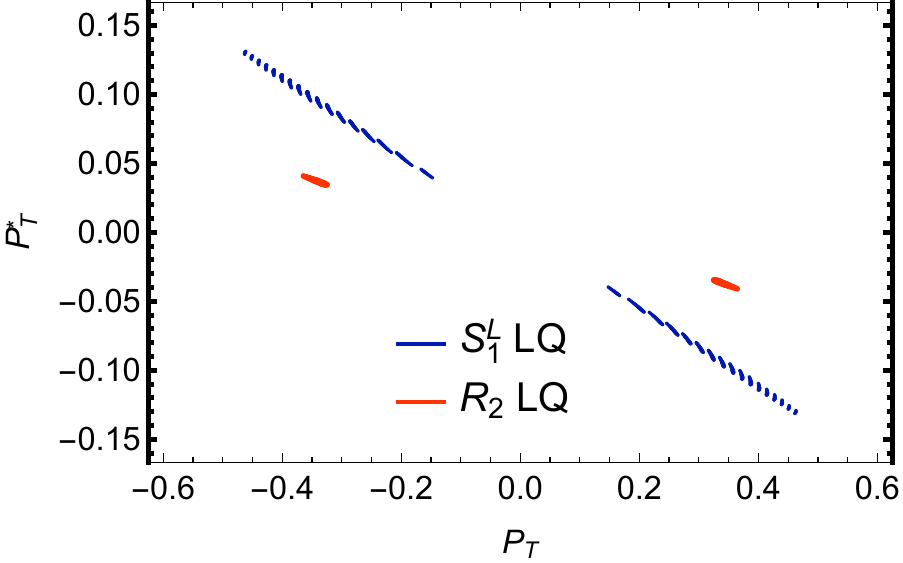}
\includegraphics[scale=0.3]{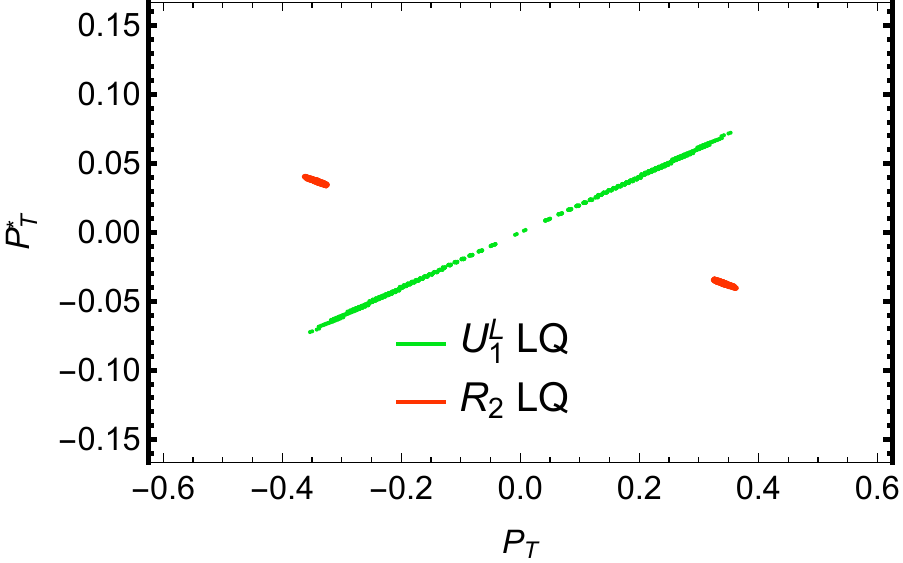}
}
\caption{The $\mathcal{P}_T$ and ${\mathcal P}_T^{*}$ observables for the points from Fig.~\ref{fig:scan_worst-case} that are less than $1\sigma$ apart in our estimation. These figures indicate that the CP-odd asymmetries $\mathcal{P}_T$ and ${\mathcal P}_T^{(*)}$ may be useful for further distinguishing the $R_2$, $U_1$, and $S_1$ leptoquark models coupled to LH neutrinos; however, the fact that they cross at the origin in the left figure also indicates that these asymmetries cannot resolve the difference in all cases.  } 
\label{fig:PT_scan1}
\end{figure}

These benchmark points illustrate that the six CP-even asymmetry observables are not enough to completely break the degeneracy between the $S_1$, $U_1$, and $R_2$ LQs when they are coupled to LH neutrinos. However, we still have a pair of observables at our disposal that could distinguish these models: $\mathcal{P}_T$ and $\mathcal{P}_T^*$. After our $\chi^2$ estimation with all eight CP-even observables, we keep the points from pairs of these LQ models that are less than $1\sigma$ apart from each other and study their contribution to the CP-odd observables $\mathcal{P}^{(*)}_T$. The results are depicted in Fig.~\ref{fig:PT_scan1}. Given a reasonable precision, say $\delta \mathcal{P}_T^{(*)}\sim0.1$, these observables are able to resolve the different degenerate models, apart from the special case where the models are CP even.

To recap, in this scenario, measurement of the CP-even asymmetry observables at Belle~II, for which theoretical proposals exist \cite{Alonso:2017ktd}, can easily discern the models of different types of neutrinos. Models with the same type of neutrinos can be distinguished in many (but not all) cases using the same CP-even measurements or with the additional measurement of CP-odd polarization asymmetries $\mathcal{P}_T^{(*)}$.

\section{Conclusion}
\label{sec:conclusion}

In this work we studied various $\tau$ asymmetry observables that can potentially be measured at Belle II and that could help to resolve the BSM origin of  the long-standing $R_{D^{(*)}}$ anomalies.
In \eqref{eq:numericRDAFB} and \eqref{eq:numericPtau}--\eqref{eq:numericPT}, we reported numerical formulas for the $\tau$ forward-backward asymmetry ${\mathcal A}_{FB}^{(*)}$ and polarization asymmetries $P_\tau^{(*)}$, $P_\perp^{(*)}$, and $P_T^{(*)}$, as a function of all relevant dimension 6 Wilson coefficients (including those for RH neutrinos). The analytic formulas from which our numerical results are derived are included in App.~\ref{sec:analytics}. While similar analytic formulas existed in the literature previously, here we report the contribution of the massless RH neutrinos as well. 

We also catalogued all the simplified models involving both LH and RH neutrinos that explain the $R_{D^{(*)}}$ anomalies and are not ruled out by the severe $Br \left( B_c \rightarrow \tau \nu \right)$ and $b\rightarrow s \nu \nu$ constraints, see Tab.~\ref{tab:operators}. We then showed that, using the CP-even asymmetry observables ${\mathcal A}_{FB}^{(*)}$, $P_\tau^{(*)}$, $P_\perp^{(*)}$ for which proposed measurement methods exist, it is possible to tell apart solutions with different types of neutrinos (SM LH vs.\ RH sterile ones) from one another, see Fig.~\ref{fig:scan_optimistic} and Fig.~\ref{fig:scan_worst-case}. In most instances, it is even possible to tell apart different mediators with the same neutrino chirality. The most useful observables for this purpose were $P_\tau^{(*)}$, followed by $P_\perp^{(*)}$ and ${\mathcal A}_{FB}^{(*)}$.

In some of the most difficult cases, the CP-even asymmetries are not enough. Here we show that the information carried in the CP-odd asymmetries $\mathcal{P}_T^{(*)}$ plays a further, crucial role in distinguishing different models. As these observables do not yet have a fully-developed experimental strategy, our results provide a strong motivation to construct one.

Our ability to distinguish between different BSM models for the  $R_{D^{(*)}}$  anomalies depends on what Belle~II actually measures for $R_{D^{(*)}}$. If Belle~II measures the $R_{D^{(*)}}$ ratios near the present values with much smaller error bars, then this measurement alone will greatly reduce the number of viable new physics models with either left- or right-handed neutrinos, compared to the present situation. In this case, it will be relatively straightforward to distinguish different models from one another using asymmetry observables. If instead, Belle~II finds $R_{D^{(*)}}$ ratios which are closer to the Standard Model prediction, while still constituting a $5\sigma$ discovery of new physics, then more models remain viable, and distinguishing between them becomes more difficult. However, in either scenario, we show that it is at least possible to distinguish between models with LH neutrinos and models with RH neutrinos through measurement of CP-even asymmetry observables.

One caveat in our work is the lack of a study of the Belle~II sensitivity to the $\tau$ asymmetries in $\bar{B}\rightarrow D^* \tau \nu$ decays. We simply assumed these observables can be measured with the same projected precision as the ones from the decay into $D$ mesons quoted in \cite{Alonso:2017ktd}. A similar, dedicated study of how to measure $\tau$  asymmetries at Belle~II in the $\bar{B}\rightarrow D^* \tau \nu$ mode is very well motivated. 

Different asymmetry observables studied here are integrals over an angle and $q^2$ of double-differential distributions (see the derivations in App.~\ref{sec:analytics}). While we have studied the angular dependence using the asymmetries, the $q^2$ dependence could in principle be used as well. Whether the $q^2$ distributions of all the $\tau$ asymmetries can be measured at Belle~II, and whether they are useful for distinguishing between different models, are interesting questions for future study.

It is also important to think about other angular measurements and their ability to tell different models apart. In particular, the $D^*$ polarization may be able to differentiate between various new physics explanations, see for example \cite{Fajfer:2012vx,Duraisamy:2013kcw,Becirevic:2016hea,Datta:2012qk,Alok:2016qyh}. However there is currently no published study of how feasible it is to measure the $D^*$ polarization at Belle~II.

Additionally, although we have focused on measuring these observables at Belle~II in this paper,  one can also consider the possibilities of doing this at LHCb. There are various reasons that suggest that LHCb will have significant difficulty in measuring these quantities with reasonable precision -- in particular lack of knowledge of the initial rest frame and generally higher background. However, it is conceivable that high statistics at LHCb and lower-background decay channels like $\tau \rightarrow l \nu \nu$ may be leveraged to obtain comparable precision in the measurement of these angular asymmetries. 

Finally, it would be interesting to consider complementary approaches to distinguishing different mediators from one another. For instance, in our study the most difficult cases generally boiled down to distinguishing various leptoquark models (e.g.~$S_1$ and $U_1$) from each other. Even though the current collider bounds on these are not severe enough to constrain these models, future direct searches would provide a complementary strategy to distinguish them, since the $S_1$ and $U_1$ leptoquarks decay differently \cite{Angelescu:2018tyl}.

Overall, this is an exciting time for $B$ physics and the $R_{D^{(*)}}$ anomalies. With Belle II coming online in the very near future, we will soon know if these anomalies are due to new physics or not. If they are due to new physics, it will be crucial to pin down the precise BSM origin of the anomalies. The results presented here are meant to be a significant step in this direction.

\section*{Acknowledgments}

We thank Diptimoy Ghosh, Anna Hallin, Roni Harnik, Mikhail Ivanov, J{\"u}rgen K{\"o}rner, Satoshi Mishima, Yuichiro Nakai, Dean Robinson, Hirose Shigeki, Olcyr Sumensari, Scott Thomas, Ryoutaro Watanabe, and Jure Zupan for enlightening discussions. The work of PA and DS is supported by DOE grant DOE-SC0010008. MRB~is supported by DOE grant DE-SC0017811. PA thanks the Kavli Institute for Theoretical Physics for the award of a graduate visiting fellowship, provided through Simons Foundation Grant No.~216179 and Gordon and Betty Moore Foundation Grant No.~4310. This research was supported in part by the National Science Foundation under Grant No.~NSF PHY17-48958.  DS is grateful to the Aspen Center for Physics, which is supported by National Science Foundation grant PHY-1607611, where this work was performed in part.

\appendix

\section{Leptonic and Hadronic Functions}
\label{sec:hadronicfuncs}

In \cite{Tanaka:2012nw,Sakaki:2013bfa}, the physics of the leptonic and the hadronic side of the processes in the $R_{D^{(*)}}$ anomalies are factorized and the relevant matrix elements are calculated. We use the leptonic and the hadronic matrix elements reported therein in our work. However, as we are working with right-handed neutrinos, one needs to calculate a few more matrix elements. In this appendix we report the new leptonic matrix elements involving right-handed neutrinos, and the new hadronic matrix element with tensor current.

The leptonic matrix elements are defined as
\begin{eqnarray}
\label{eq:lepSr}
Lr^{\lambda_\tau} =&& 2  \langle		\tau (p_\tau,\lambda_\tau) \bar{\nu} |		\bar{\tau}  P_R \nu		|		0		\rangle, \\
\label{eq:lepVr}
Lr_{\bar{\lambda}}^{\lambda_\tau} =&& 2 \epsilon_\mu (\bar{\lambda}) \langle		\tau (p_\tau,\lambda_\tau) \bar{\nu} |		\bar{\tau} \gamma^\mu P_R \nu		|		0		\rangle, \\
\label{eq:lepTr}
Lr_{\bar{\lambda}\bar{\lambda}'}^{\lambda_\tau} =&& -2i \epsilon_\mu (\bar{\lambda}) \epsilon_\nu (\bar{\lambda}') \langle		\tau (p_\tau,\lambda_\tau) \bar{\nu} |		\bar{\tau} \sigma^{\mu\nu} P_R \nu		|		0		\rangle ,
\end{eqnarray}
where $\bar{\lambda}$ ($\lambda_\tau$) denotes the polarization of the mediator ($\tau$ lepton). We use the same convention for the $\epsilon$ as \cite{Hagiwara:1989cu}. Explicitly carrying out the calculation, we find the following results for different polarizations. 
\begin{eqnarray}
\label{eq:scalarr+} 
Lr^+ =&&  0, \\
\label{eq:scalarr-} 
Lr^- =&& 2\sqrt{q^2} v.
\end{eqnarray}
\begin{eqnarray}
\label{eq:right+pm}
Lr^+_\pm =&&  -\sqrt{2} \sqrt{q^2}  v (1 \mp \cos \theta)  \\
\label{eq:right+0}
Lr^+_0 =&& - 2 \sqrt{q^2}  v \sin \theta \\
\label{eq:right+s}
Lr^+_t =&& 0 \\
\label{eq:right-pm}
Lr^-_\pm =&& \mp \sqrt{2} m_\tau v \sin \theta \\
\label{eq:right-0}
Lr^-_0 =&& - 2 m_\tau v \cos \theta \\
\label{eq:right-s}
Lr^-_t =&& 2 m_\tau v,
\end{eqnarray}
\begin{eqnarray}
\label{eq:tensR1}
Lr^\pm_{\lambda \lambda}=&& 0,\\
\label{eq:tensR2}
Lr^\pm_{\lambda \lambda'}=&& -Lr^\pm_{\lambda' \lambda},\\
\label{eq:tensR3}
Lr^+_{0 \pm}=&&\mp \sqrt{2} m_\tau v (1 \mp \cos \theta)		 	,\\
\label{eq:tensR4}
Lr^+_{+-} = &&- Lr^+_{t0}=2 m_\tau v \sin \theta		 	,\\
\label{eq:tensR5}
Lr^+_{\pm t} =&& \sqrt{2} m_\tau v (1 \mp \cos \theta)		 	,\\
\label{eq:tensR6}
Lr^-_{0 \pm}=&& - \sqrt{2} \sqrt{q^2} v \sin \theta		 	,\\
\label{eq:tensR7}
Lr^-_{+-} =&& - Lr^-_{t0}=2 \sqrt{q^2} v \cos \theta		 	,\\
\label{eq:tensR8}
Lr^-_{\pm t} =&&	  \pm	\sqrt{2} \sqrt{q^2}v \sin \theta	 	,
\end{eqnarray}
where $\theta$ is again the angle between the $\tau$ lepton and the $D^{(*)}$ in the leptonic system rest-frame, see Fig.~\ref{fig:kinematics}, and $v=\sqrt{1- m_\tau^2/q^2}$. The subscript $t$ refers to the fourth polarization of a virtual $W$. These leptonic functions are related to the ones involving the LH neutrinos in \cite{Tanaka:2012nw} through parity transformation.

For the hadronic side of the matrix element, we use the notation from \cite{Sakaki:2013bfa} and define
\begin{eqnarray}
\label{eq:RDhad}
H_{V,0}^s (q^2)\equiv && H_{V1,0}^s (q^2)=H_{V2,0}^s (q^2),\nonumber\\
H_{V,t}^s (q^2)\equiv && H_{V1,t}^s (q^2)=H_{V2,t}^s (q^2),\nonumber\\
H_{S}^s (q^2)\equiv && H_{S1}^s (q^2)=H_{S2}^s (q^2),\\
H_{T}^s (q^2)\equiv && H_{T,+-}^s (q^2)=H_{T,0t}^s (q^2) = - H_{T2,+-}^s (q^2)=H_{T2,0t}^s (q^2).\nonumber
\end{eqnarray}
\begin{eqnarray}
\label{eq:RDshad}
H_{V,\pm} (q^2)\equiv && H_{V1,\pm}^\pm (q^2)=-H_{V2,\mp}^\mp (q^2),\nonumber\\
H_{V,0} (q^2)\equiv && H_{V1,0}^0 (q^2)=-H_{V2,0}^0 (q^2),\nonumber\\
H_{V,t} (q^2)\equiv && H_{V1,t}^0 (q^2)=-H_{V2,t}^0 (q^2),\nonumber\\
H_{S} (q^2) \equiv && H_{S1}^0 (q^2)=-H_{S2}^0 (q^2),\\
H_{T,\pm} (q^2) \equiv  && H^\pm_{T,\pm 0} = \pm H_{T,\pm t}^\pm (q^2),\nonumber\\
H_{T2,\pm} (q^2) \equiv &&   H^\pm_{T2,\pm 0} = \mp H_{T2,\pm t}^\pm (q^2),\nonumber\\
H_{T,0} (q^2) \equiv && H^0_{T,+-}(q^2)=H_{T,0 t}^0 (q^2) = H^0_{T2,+-}(q^2)= - H_{T2,0 t}^0 (q^2).\nonumber
\end{eqnarray}
The hadronic functions in \eqref{eq:RDhad} correspond to the $\bar{B}\rightarrow D \tau \nu$ decay, while those in \eqref{eq:RDshad} correspond to the $\bar{B}\rightarrow D^* \tau \nu$ decay. The superscripts $0$ and $\pm$ in \eqref{eq:RDshad} stand for the $D^*$ polarization, while the subscripts $0$, $\pm$, and $t$ refer to the virtual mediator polarization. The subscripts $S$, $V$, and $T$ refer to the scalar, vector, and tensor currents, respectively.

In the $\bar{B}\rightarrow D \tau \nu$ hadronic functions, we use the same form factors as in \cite{Bardhan:2016uhr} (derived from the available lattice results \cite{Lattice:2015rga} and from \cite{Melikhov:2000yu}); for the $\bar{B}\rightarrow D^* \tau \nu$ decay, following \cite{Bardhan:2016uhr,Sakaki:2013bfa}, we use the heavy quark effective theory form factors based on \cite{Caprini:1997mu}.

When working with the right-handed neutrino models for $R_{D^{(*)}}$, there is one new hadronic function, namely
\begin{equation}
\langle		D^{(*)}	|		\bar{c} \sigma^{\mu\nu} (1+\gamma^5) b		| B			\rangle,
\label{eq:hadfuncT}
\end{equation}
where $\sigma^{\mu\nu} = i/2 [\gamma^\mu,\gamma^\nu]$. To calculate this matrix element, we can simply borrow the results in \cite{Sakaki:2013bfa} for the hadronic side of the operator $\mathcal{O}^T_{LL}$ ($\bar{c} \sigma^{\mu\nu} (1-\gamma^5) b$ operator) and merely flip the sign of the axial current. The resulting hadronic functions, denoted by $H_{T2}$, will be 
\begin{equation}
\label{eq:hadronsrd1T2}
H_{T2,+-}^s (q^2)= - H_{T2,0t}^s (q^2) = - H_{T,+-}^s (q^2),
\end{equation}
where the superscript $s$ indicates that these functions are corresponding to $D$ meson and $H_{T}$ functions are defined in \cite{Sakaki:2013bfa}, and 
\begin{eqnarray}
\label{eq:hadronsrds2T2}
H_{T2,+-}^0 (q^2)&= & - H_{T2,0 t}^0 (q^2) = H_{T,+-}^0 (q^2), \nonumber \\
\label{eq:hadronsrds1T2}
H_{T2,\pm 0}^\pm (q^2)&= &\mp H_{T2,\pm t}^\pm (q^2) = \frac{\sqrt{m_B m_{D^*}}}{\sqrt{q^2}} A_1(w) \left( \pm (m_b-m_c) (w+1) - (m_b+m_c) \sqrt{w^2-1} R_1 (w)		\right), \nonumber \\
\label{eq:hadronsrds3T2}
H_{T2,\lambda_1 \lambda_2}^{\lambda_M} (q^2)&=& - H_{T2,\lambda_2 \lambda_1}^0 (q^2), 
\end{eqnarray}
for $D^*$ where again the $H_{T}$ functions and the form factors $A_1(w)$ and $R_1(w)$ are defined in \cite{Tanaka:2012nw,Sakaki:2013bfa}, $m_M$ is the final meson (here $D^*$) mass, and 
\begin{equation}
w = \frac{m^2_B+m_M^2-q^2}{2m_Mm_B}.
\label{eq:wmesons}
\end{equation}

\section{Analytic Expressions for the Observables}
\label{sec:analytics}

In order to get an expression for the polarization asymmetries, we write the total decay rate with the spin of the final state $\tau$ lepton in the arbitrary direction $\hat{s}$ as \cite{Tanaka:1994ay} 
\begin{equation}
d\Gamma^{(*)} (\hat{s}) = \frac{1}{2} \left( 	d\Gamma^{(*)}_{\rm tot} + \left( d\Gamma^{(*)}_\tau \hat{e}_\tau + d\Gamma^{(*)}_\perp \hat{e}_\perp  + d\Gamma^{(*)}_T \hat{e}_T			\right)\cdot \hat{s}		\right),
\label{eq:taupolzM}
\end{equation}
where we have suppressed all other final state indices, e.g. $D^*$ polarization, and
\begin{eqnarray}
d\Gamma^{(*)}_{\rm tot} &=& \frac{1}{2m_B} \left(	|\mathcal{M}_+|^2 + |\mathcal{M}_-|^2	\right) d\Phi, \nonumber \\
\label{eq:polzMs}
d\Gamma^{(*)}_\tau &=& \frac{1}{2m_B} \left(	|\mathcal{M}_+|^2 - |\mathcal{M}_-|^2	\right) d\Phi, \\
d\Gamma^{(*)}_\perp &=& \frac{1}{2m_B} 2 \mathcal{R}e\left(	\mathcal{M}_+^\dagger	 \mathcal{M}_- \right) d\Phi, \nonumber \\
d\Gamma^{(*)}_T &=& \frac{1}{2m_B} 2 \mathcal{I}m\left(	\mathcal{M}_+^\dagger \mathcal{M}_-	\right) d\Phi,\nonumber 
\end{eqnarray}
where $\mathcal{M}_\pm$ are the corresponding matrix elements with $\pm$ $\tau$ helicity. The phase space element $d\Phi$ is given by
\begin{equation}
d\Phi = \frac{\sqrt{\left(  (m_B + m_M)^2 - q^2       \right)   \left(   (m_B - m_M)^2 - q^2        \right)}}{256 \pi^3 m_B^2} \left(		1-\frac{m_\tau^2}{q^2}			\right)^2 dq^2 d\cos \theta,
\label{eq:phasespace}
\end{equation}
with $m_M$ being the final meson ($D^{(*)}$) mass, $q^2$ being the four-momentum transferred to the leptonic side, and $\theta$ being the angle between the $\tau$ momentum and the final meson $M$ in the $q^2$ frame. Using \eqref{eq:polzMs} in \eqref{eq:defPt} we find an expression for the integrated asymmetries in every direction
\begin{equation}
\mathcal{P}_{x}^{(*)} = \frac{1}{\Gamma_{\mathrm{tot}}^{(*)} } \int d\Gamma_x^{(*)} ,
\label{eq:defPt2}
\end{equation}
where $x=\tau,\perp,T$. Similarly, the forward-backward symmetry defined in \eqref{eq:defAFB} can be written as
\begin{equation}
\mathcal{A}_{FB}^{(*)} = \frac{1}{\Gamma_{\mathrm{tot}}^{(*)}} \int_{q^2} \left(	-\int_{\cos\left(\theta \right)=0}^{\cos\left(\theta \right)=1} + \int_{\cos\left(\theta \right)=-1}^{\cos\left(\theta \right)=0}	\right) d\Gamma_{\mathrm{tot}}^{(*)}
\label{eq:defAFBappx}
\end{equation}

In this appendix we use \eqref{eq:defPt2} for the polarization asymmetries, as well as \eqref{eq:defAFBappx} for the forward-backward asymmetry $\mathcal{A}_{FB}^{(*)}$, to find analytic formulas for different decay rates used in Sec.~\ref{sec:obsevables}. As indicated in the previous appendix, we use the convention and the notation in \cite{Sakaki:2013bfa} for the hadronic functions. 

We start with $\Gamma^{(*)}_{\mathrm{tot}}$ and $\Gamma^{(*)}_{\tau}$. For the LH neutrinos contribution to the $\bar{B} \rightarrow D \tau \nu$ we have
\begin{eqnarray}
\frac{d\Gamma_\mathrm{tot}}{dq^2} = && \frac{G_F^2 V_{cb}^2}{192 m_B^3 \pi^3 q^2}   \sqrt{\left(	(m_B-m_D)^2-q^2 \right) \left(	(m_B+m_D)^2-q^2 \right)  }  (m_\tau^2 - q^2)^2 \nonumber \\
 && \left\lbrace	 |1+C^V_{LL}+C^V_{RL}|^2 \left[	(H^{s}_{V,0})^2  \left(\frac{m_\tau^2}{2q^2} + 1 \right) + \frac{3 m_\tau^2}{2q^2} (H^s_{V,t})^2	 	\right] \right.  \nonumber  \\
\label{eq:analyticRDLPtaupl}
&& + \frac{3}{2} (H^s_S)^2 |C^S_{RL}+C^S_{LL}|^2 +8 |C^T_{LL}|^2  (H^s_T)^2 \left(	1+ \frac{2m_\tau^2}{q^2}	\right) \nonumber\\
&& + 3 \mathcal{R}e\left[	(1+C^V_{LL}+C^V_{RL}) (C^{S}_{RL}+C^{S}_{LL})^{*}	\right] \frac{m_\tau}{\sqrt{q^2}} H_S^s H^s_{V,t} \nonumber \\
&& - \left. 12 \mathcal{R}e \left[	(1+C^V_{LL}+C^V_{RL}) (C^T_{LL})^{*}		\right] \frac{m_\tau}{\sqrt{q^2}} H^s_T H^s_{V,0} \nonumber \right\rbrace , \\
\\
\frac{d\Gamma_\tau}{dq^2} = && \frac{G_F^2 V_{cb}^2}{192 m_B^3 \pi^3 q^2}   \sqrt{\left(	(m_B-m_D)^2-q^2 \right) \left(	(m_B+m_D)^2-q^2 \right)  }  (m_\tau^2 - q^2)^2 \nonumber \\
 && \left\lbrace	 |1+C^V_{LL}+C^V_{RL}|^2 \left[	(H^{s}_{V,0})^2  \left(\frac{m_\tau^2}{2q^2} - 1 \right) + \frac{3 m_\tau^2}{2q^2} (H^s_{V,t})^2	 	\right] \right.  \nonumber  \\
&& + \frac{3}{2} (H^s_S)^2 |C^S_{RL}+C^S_{LL}|^2 +8 |C^T_{LL}|^2  (H^s_T)^2 \left(	1- \frac{2m_\tau^2}{q^2}	\right) \nonumber\\
&& + 3 \mathcal{R}e\left[	(1+C^V_{LL}+C^V_{RL}) (C^{S}_{RL}+C^{S}_{LL})^{*}	\right] \frac{m_\tau}{\sqrt{q^2}} H_S^s H^s_{V,t} \nonumber \\
&& + \left. 4 \mathcal{R}e \left[	(1+C^V_{LL}+C^V_{RL}) (C^T_{LL})^{*}		\right] \frac{m_\tau}{\sqrt{q^2}} H^s_T H^s_{V,0} \nonumber \right\rbrace .
\end{eqnarray}

Similarly, the contribution of the RH neutrinos to these rates are 
\begin{eqnarray}
\frac{d\Gamma_{\mathrm{tot}}}{dq^2} = && \frac{G_F^2 V_{cb}^2}{192 m_B^3 \pi^3 q^2}   \sqrt{\left(	(m_B-m_D)^2-q^2 \right) \left(	(m_B+m_D)^2-q^2 \right)  }  (m_\tau^2 - q^2)^2  \nonumber \\
&& \left\lbrace |C^V_{LR}+C^V_{RR}|^2 \left(	 (H_{V,0}^s)^2 \left( \frac{m_\tau^2}{2q^2} +1 \right)+ \frac{3m_\tau^2}{2q^2} (H_{V,t}^s)^2	\right)	 \right.  \nonumber  \\
&& +  \frac{3}{2} |C^S_{RR}+C^S_{LR}|^2 (H_{S}^{s})^2 + 8 |C^T_{RR}|^2  (H^s_{T})^2 \left(	1 + \frac{2m_\tau^2}{q^2} 	\right) \nonumber \\
&& + 3 \mathcal{R}e \left[	(C^S_{RR}+C^S_{LR}) (C^V_{LR}+C^V_{RR})^* 	\right] \frac{m_\tau}{\sqrt{q^2}} H_{S}^s  H_{V,t}^s \nonumber \\
&& - \left. 12 \mathcal{R}e\left[	(C^V_{RR}+C^V_{LR}) (C^T_{RR})^*	\right] \frac{m_\tau}{\sqrt{q^2}} H_{T}^s H^s_{V,0}\right\rbrace  \nonumber  \\
\label{eq:analyticRDRPtaupl}
\\
\frac{d\Gamma_\tau}{dq^2} = && \frac{G_F^2 V_{cb}^2}{192 m_B^3 \pi^3 q^2}   \sqrt{\left(	(m_B-m_D)^2-q^2 \right) \left(	(m_B+m_D)^2-q^2 \right) }  (m_\tau^2 - q^2)^2  \nonumber \\
&& \left\lbrace |C^V_{LR}+C^V_{RR}|^2 \left(	 (H_{V,0}^s)^2 \left( - \frac{m_\tau^2}{2q^2} +1 \right) - \frac{3m_\tau^2}{2q^2} (H_{V,t}^s)^2	\right)	 \right.  \nonumber  \\
&& -  \frac{3}{2} |C^S_{RR}+C^S_{LR}|^2 (H_{S}^{s})^2 + 8 |C^T_{RR}|^2  (H^s_{T})^2 \left(	- 1 + \frac{2m_\tau^2}{q^2} 	\right) \nonumber \\
&& - 3 \mathcal{R}e \left[	(C^S_{RR}+C^S_{LR}) (C^V_{LR}+C^V_{RR})^* 	\right] \frac{m_\tau}{\sqrt{q^2}} H_{S}^s  H_{V,t}^s \nonumber \\
&& - \left. 4 \mathcal{R}e\left[	(C^V_{RR}+C^V_{LR}) (C^T_{RR})^*	\right] \frac{m_\tau}{\sqrt{q^2}} H_{T}^s H^s_{V,0}  \right\rbrace \nonumber .
\end{eqnarray}
The dependence of all the hadronic functions $H$ on $q^2$ is implicit. These equations can be used to calculate the contribution of each type of neutrinos to $R_D$ and $\mathcal{P}_\tau$.

For the LH neutrinos contribution to $\bar{B} \rightarrow D^* \tau \nu$ we have
\begin{eqnarray}
\frac{d\Gamma_\mathrm{tot}^*}{dq^2} = && \frac{G_F^2 V_{cb}^2}{192 m_B^3 \pi^3 q^2}   \sqrt{\left(	(m_B-m_{D^*})^2-q^2 \right) \left(	(m_B+m_{D^*})^2-q^2 \right)  }  (m_\tau^2 - q^2)^2\nonumber \\
 && \left\lbrace	 (|1+C^V_{LL}|^2+|C^V_{RL}|^2) \left[	(H^{2}_{V,+}+H^{2}_{V,-}+H^{2}_{V,0})  \left( \frac{m_\tau^2}{2q^2} + 1\right) + \frac{3 m_\tau^2}{2q^2} H^2_{V,t}	\right] \right.  \nonumber  \\
 &&	-2\mathcal{R}e \left[ (1+C^V_{LL})(C^{V}_{RL})^{*} \right]\left[	(2H_{V,+}H_{V,-}+H^{2}_{V,0})  \left( \frac{m_\tau^2}{2q^2} + 1\right)+ \frac{3 m_\tau^2}{2q^2} H^2_{V,t}	\right]  \nonumber  \\
&& + \frac{3}{2} (H_S)^2 |C^S_{RL}-C^S_{LL}|^2 +8 |C^T_{LL}|^2 \left(	1+ \frac{2m_\tau^2}{q^2}		\right) (H^2_{T,+}+H^2_{T,-}+H^2_{T,0}) \nonumber \\
&& + 3 \mathcal{R}e\left[	(1+C^V_{LL}-C^V_{RL}) (C^{S}_{RL}-C^S_{LL})^{*}	\right] \frac{m_\tau}{\sqrt{q^2}} H_S H_{V,t} \nonumber \\
&& - 12 \mathcal{R}e \left[	(1+C^V_{LL}) (C^T_{LL})^{*}		\right] \frac{m_\tau}{\sqrt{q^2}} (H_{T,0} H_{V,0} + H_{T,+} H_{V,+}  - H_{T,-} H_{V,-} ) \nonumber \\
&&  + \left. 12 \mathcal{R}e \left[	C^V_{RL} (C^T_{LL})^{*}		\right] \frac{m_\tau}{\sqrt{q^2}} (H_{T,0} H_{V,0} + H_{T,+} H_{V,-}  - H_{T,-} H_{V,+} )  \right\rbrace, \nonumber \\
\label{eq:analyticRDsLPtaupl}
\\
\frac{d\Gamma_\tau^*}{dq^2} = && \frac{G_F^2 V_{cb}^2}{192 m_B^3 \pi^3 q^2}   \sqrt{\left(	(m_B-m_{D^*})^2-q^2 \right) \left(	(m_B+m_{D^*})^2-q^2 \right)  }  (m_\tau^2 - q^2)^2\nonumber \\
 && \left\lbrace	 (|1+C^V_{LL}|^2+|C^V_{RL}|^2) \left[	(H^{2}_{V,+}+H^{2}_{V,-}+H^{2}_{V,0})  \left( \frac{m_\tau^2}{2q^2} - 1\right) + \frac{3 m_\tau^2}{2q^2} H^2_{V,t}	\right] \right.  \nonumber  \\
 &&	-2\mathcal{R}e \left[ (1+C^V_{LL})(C^{V}_{RL})^{*} \right]\left[	(2H_{V,+}H_{V,-}+H^{2}_{V,0})  \left( \frac{m_\tau^2}{2q^2} - 1\right)+ \frac{3 m_\tau^2}{2q^2} H^2_{V,t}	\right]  \nonumber  \\
&& + \frac{3}{2} (H_S)^2 |C^S_{RL}-C^S_{LL}|^2 +8 |C^T_{LL}|^2 \left(	1- \frac{2m_\tau^2}{q^2}		\right) (H^2_{T,+}+H^2_{T,-}+H^2_{T,0}) \nonumber \\
&& + 3 \mathcal{R}e\left[	(1+C^V_{LL}-C^V_{RL}) (C^{S}_{RL}-C^S_{LL})^{*}	\right] \frac{m_\tau}{\sqrt{q^2}} H_S H_{V,t} \nonumber \\
&& + 4 \mathcal{R}e \left[	(1+C^V_{LL}) (C^T_{LL})^{*}		\right] \frac{m_\tau}{\sqrt{q^2}} (H_{T,0} H_{V,0} + H_{T,+} H_{V,+}  - H_{T,-} H_{V,-} ) \nonumber \\
&&  - \left. 4 \mathcal{R}e \left[	C^V_{RL} (C^T_{LL})^{*}		\right] \frac{m_\tau}{\sqrt{q^2}} (H_{T,0} H_{V,0} + H_{T,+} H_{V,-}  - H_{T,-} H_{V,+} )  \right\rbrace \nonumber .
\end{eqnarray}

The corresponding decay rates with RH neutrinos instead are 
\begin{eqnarray}
\frac{d\Gamma_\mathrm{tot}^*}{dq^2} = && \frac{G_F^2 V_{cb}^2}{192 m_B^3 \pi^3 q^2}   \sqrt{\left(	(m_B-m_{D^*})^2-q^2 \right) \left(	(m_B+m_{D^*})^2-q^2 \right)  }  (m_\tau^2 - q^2)^2 \nonumber \\
 && \left\lbrace	 (|C^V_{LR}|^2 + |C^V_{RR}|^2) \left( \frac{m_\tau^2}{2q^2} + 1 \right) \left( (H_{V,-})^2 +(H_{V,+})^2 \right) \right.  \nonumber  \\
&& + |C^V_{LR} - C^V_{RR}|^2 \left(		(H_{V,0})^2 \left(	\frac{m_\tau^2}{2q^2}  + 1	\right)  +   (H_{V,t})^2	\frac{3m_\tau^2}{2q^2}	\right) \nonumber \\ 
&& -	4\mathcal{R}e \left[ C^V_{LR}(C^{V}_{RR})^{*} \right]  \left( \frac{m_\tau^2}{2q^2} + 1 \right)  H_{V,+} H_{V,-}	+ \frac{3}{2}  |C^S_{RR}-C^S_{LR}|^2  (H_{S})^2 \nonumber  \\
&& + 3 \mathcal{R}e\left[	(C^V_{RR}-C^V_{LR}) (C^{S}_{LR}-C^{S}_{RR})^{*}	\right] \frac{m_\tau}{\sqrt{q^2}} H_{S} H_{V,t} \nonumber \\
&& + 8  |C^T_{RR}|^2 \left(	1  + \frac{2m_\tau^2}{q^2} 	\right) \left(	 (H_{T2,-})^2  + (H_{T,0})^2 +   (H_{T2,+})^2    \right) \nonumber \\
&& - 12 \mathcal{R}e \left[	C^V_{LR} (C^{T}_{RR})^*		\right] \frac{m_\tau}{\sqrt{q^2}} (H_{T2,-} H_{V,-} - H_{T,0} H_{V,0}  - H_{T2,+} H_{V,+} )  \nonumber \\
&& + \left. 12 \mathcal{R}e \left[	C^V_{RR} (C^{T}_{RR})^*		\right] \frac{m_\tau}{\sqrt{q^2}} (H_{T2,-} H_{V,+} - H_{T,0} H_{V,0}  - H_{T2,+} H_{V,-} )  \nonumber \right\rbrace \\
\label{eq:analyticRDsRPtaupl}
\\
\frac{d\Gamma_\tau^*}{dq^2} = && \frac{G_F^2 V_{cb}^2}{192 m_B^3 \pi^3 q^2}   \sqrt{\left(	(m_B-m_{D^*})^2-q^2 \right) \left(	(m_B+m_{D^*})^2-q^2 \right)  }  (m_\tau^2 - q^2)^2 \nonumber \\
 && \left\lbrace	 (|C^V_{LR}|^2 + |C^V_{RR}|^2) \left( - \frac{m_\tau^2}{2q^2} + 1 \right) \left( (H_{V,-})^2 +(H_{V,+})^2 \right) \right.  \nonumber  \\
&& + |C^V_{LR} - C^V_{RR}|^2 \left(		(H_{V,0})^2 \left(	- \frac{m_\tau^2}{2q^2}  + 1	\right)  -   (H_{V,t})^2	\frac{3m_\tau^2}{2q^2}	\right) \nonumber \\ 
&& -	4\mathcal{R}e \left[ C^V_{LR}(C^{V}_{RR})^{*} \right]  \left( - \frac{m_\tau^2}{2q^2} + 1 \right)  H_{V,+} H_{V,-}	- \frac{3}{2}  |C^S_{RR}-C^S_{LR}|^2  (H_{S})^2 \nonumber  \\
&& - 3 \mathcal{R}e\left[	(C^V_{RR}-C^V_{LR}) (C^{S}_{LR}-C^{S}_{RR})^{*}	\right] \frac{m_\tau}{\sqrt{q^2}} H_{S} H_{V,t} \nonumber \\
&& + 8  |C^T_{RR}|^2 \left(	- 1  + \frac{2m_\tau^2}{q^2} 	\right) \left(	 (H_{T2,-})^2  + (H_{T,0})^2 +   (H_{T2,+})^2    \right) \nonumber \\
&& - 4 \mathcal{R}e \left[	C^V_{LR} (C^{T}_{RR})^*		\right] \frac{m_\tau}{\sqrt{q^2}} (H_{T2,-} H_{V,-} - H_{T,0} H_{V,0}  - H_{T2,+} H_{V,+} )  \nonumber \\
&& + \left. 4 \mathcal{R}e \left[	C^V_{RR} (C^{T}_{RR})^*		\right] \frac{m_\tau}{\sqrt{q^2}} (H_{T2,-} H_{V,+} - H_{T,0} H_{V,0}  - H_{T2,+} H_{V,-} )  \nonumber \right\rbrace .
\end{eqnarray}
We can use \eqref{eq:analyticRDsLPtaupl}--\eqref{eq:analyticRDsRPtaupl} to find the contribution of each type of neutrinos to $R_{D^*}$ and $\mathcal{P}^*_\tau$.

\begin{figure}
\includegraphics[scale=0.75]{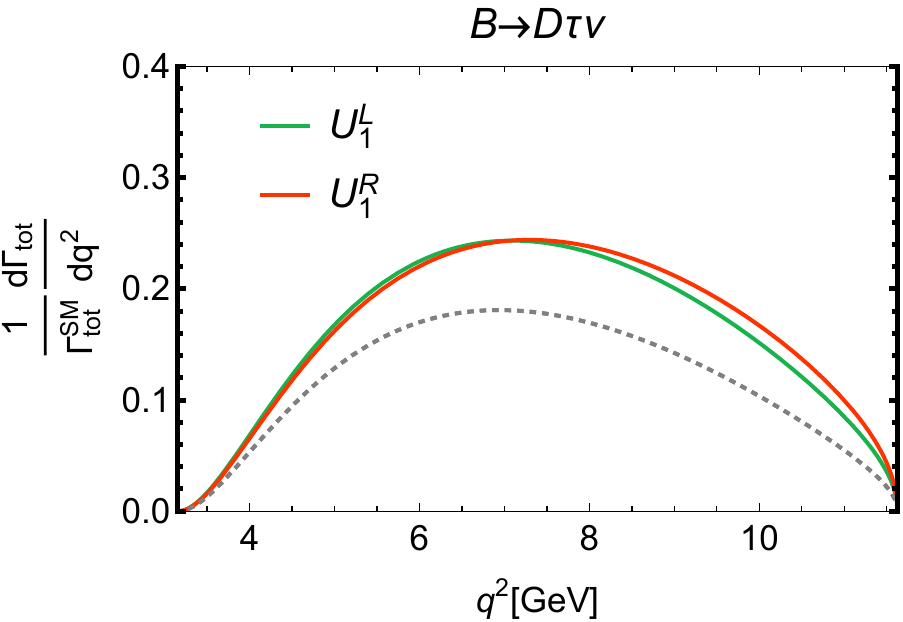}
\includegraphics[scale=0.75]{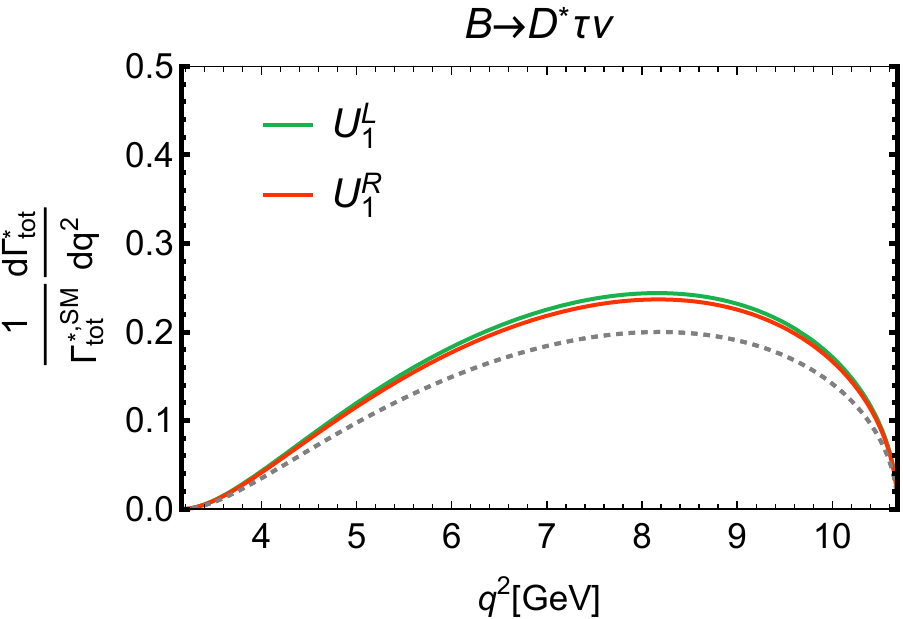}
\caption{The $q^2$ distribution for benchmark Wilson coefficients for models interacting with LH neutrinos (green curves) and those interacting with RH neutrinos (red curves). The decay rate for $\bar{B}\rightarrow D \tau \nu$ ($\bar{B}\rightarrow D^* \tau \nu$) is shown on left (right). We show the viable LQ models whose effective operators from Tab.~\ref{tab:operators} are related through the symmetry in \eqref{eq:transform}--\eqref{eq:transform2}. The dashed gray line is the SM prediction. The area under each curve is proportional to its prediction for $R_{D^{(*)}}$. Up to a rescaling factor the plots for different types of neutrinos have indistinguishable shapes.}
\label{fig:q2dist}
\end{figure}

The symmetry outlined in \eqref{eq:transform}--\eqref{eq:transform2} between RH and LH neutrino contribution is manifested in the decay rates \eqref{eq:analyticRDLPtaupl}--\eqref{eq:analyticRDsRPtaupl}. In other words, the unpolarized decay rates have the same $q^2$ dependence for either types of neutrinos. The only difference between the two cases is the irreducible SM contribution. Even considering this difference, there are scenarios with different types of neutrinos that have indistinguishable $q^2$ distributions. An example of this is illustrated in  Fig.~\ref{fig:q2dist} for $U_1$ LQ models interacting with different types of neutrinos. In this figure we normalize the differential rate to the SM total rate, so that the area under each plot is proportional to its $R_{D^{(*)}}$ prediction. 
Each curve results in an $R_{D^{(*)}}$ close to the current global averages (see the $10\sigma$ scenario in Sec.~\ref{subsec:scenarios}). These plots show that in this scenario there are benchmark points that, unlike the asymmetry observables we studied, the $q^2$ distribution of the decay rates will not be able to distinguish models with different types of neutrinos.  

For  $\Gamma_{\perp}$ and $\Gamma_{T}$ we have 
\begin{eqnarray}
\label{eq:PTPNRDanalytic}
{d\Gamma_{\perp}\over dq^2} &= & \frac{G_F^2 V_{cb}^2}{192 m_B^3 \pi^3 q^2}   \sqrt{\left(	(m_B-m_D)^2-q^2 \right) \left(	(m_B+m_D)^2-q^2 \right)  } (m_\tau^2 - q^2)^2 \mathcal{R}e \left( \Sigma \right) ,\nonumber  \\
\\
{d\Gamma_{T}\over dq^2} &= & \frac{G_F^2 V_{cb}^2}{192 m_B^3 \pi^3 q^2}   \sqrt{\left(	(m_B-m_D)^2-q^2 \right) \left(	(m_B+m_D)^2-q^2 \right)  } (m_\tau^2 - q^2)^2 \mathcal{I}m \left( \Sigma \right) \nonumber  ,
\end{eqnarray}
where 
\begin{eqnarray}
\label{eq:PTPNRDsigma}
\Sigma &=& \frac{3\pi }{4q^2}  \left[  	  \left(		 \sqrt{q^2} (C^S_{RR} + C^S_{LR}) H^s_S  	 + m_\tau  	(C^V_{LR} + C^V_{RR}) H^s_{V,t}  		\right)	\times		\right. \nonumber \\
&& \left(		(C^V_{LR} + C^V_{RR})^* \sqrt{q^2}  	H^s_{V0,0}    -4m_\tau (C^T_{RR})^*  H^s_T   	  \right) -  \\
&& \left(		(C^S_{RL} + C^S_{LL})^*	\sqrt{q^2} H^s_S  	 +   m_\tau  	(1+C^V_{LL} + C^V_{RL})^* H^s_{V,t}  	   \right)  \times\nonumber \\
&&\left.  	\left(	 	(1 +C^V_{LL} + C^V_{RL}) \sqrt{q^2}  	H^s_{V,0}    -4 m_\tau (C^T_{LL})  H^s_T   	  \right)		\right] 			\nonumber
\end{eqnarray}
Equivalently, we can write the polarization asymmetries in $\bar{B} \rightarrow D^* \tau \nu$ as
\begin{eqnarray}
\label{eq:PTPNRDsanalytic}
{d\Gamma_{\perp}^*\over dq^2} &=& \frac{G_F^2 V_{cb}^2}{192 m_B^3 \pi^3 q^2}   \sqrt{\left(	(m_B-m_{D^*})^2-q^2 \right) \left(	(m_B+m_{D^*})^2-q^2 \right)  } (m_\tau^2 - q^2)^2 \mathcal{R}e \left( \Sigma^* \right) ,\nonumber  \\
\\
{d\Gamma_{T}^* \over dq^2} &= & \frac{G_F^2 V_{cb}^2}{192 m_B^3 \pi^3 q^2}   \sqrt{\left(	(m_B-m_{D^*})^2-q^2 \right) \left(	(m_B+m_{D^*})^2-q^2 \right)  } (m_\tau^2 - q^2)^2 \mathcal{I}m \left( \Sigma^* \right) \nonumber  ,
\end{eqnarray}
where $\Sigma^*$ is given by
\begin{eqnarray}
\label{eq:PTPNRDssigma}
\Sigma^* &=& \frac{3\pi}{8q^2} \left[			- \left(		(1+C^V_{LL}) \sqrt{q^2}  H_{V,-}    - \sqrt{q^2}  C^V_{RL}  H_{V,+}  		+   4 C^T_{LL}		m_\tau H_{T,-}   	  \right) \times 		\right. \nonumber \\
&& \left( 		m_\tau (1+C^V_{LL})^* H_{V,-}     -  m_\tau  (C^V_{RL})^*	H_{V,+}  		+ 4  		\sqrt{q^2}   	(C^T_{LL})^*   H_{T,-}   		\right) \nonumber \\
&+& \left(		2\sqrt{q^2} (C^S_{LL}-C^S_{RL})^* H_S   + 		 2m_\tau 	(-1-C^V_{LL}+C^V_{RL})^* H_{V,t}   		\right) \times \nonumber \\
&& \left(			(1+C^V_{LL}-C^V_{RL}) \sqrt{q^2}  H_{V,0}    -  4  C^T_{LL}  m_\tau  H_{T,0}  		\right) \nonumber \\
&+&  \left(			\sqrt{q^2} C^V_{RL} H_{V,-}     -  \sqrt{q^2} (1+C^V_{LL})  H_{V,+}  	   +  4 m_\tau  			C^T_{LL} H_{T,+} 			\right)  \times \nonumber \\	
&&  \left(			m_\tau (C^V_{RL})^*	 H_{V,-}    	- m_\tau (1+C^V_{LL})^*		H_{V,+}     +  4   \sqrt{q^2}  (C^T_{LL})^*		H_{T,+} 			\right)  \nonumber \\	
\\
&-&  \left(		(C^V_{RR})^* \sqrt{q^2}  H^+_{V,+}    - \sqrt{q^2}  (C^V_{LR})^*  H^-_{V,-}  		+   4 (C^T_{RR})^*		m_\tau H_{T2,-}   	  \right) \times 	 \nonumber \\
&& \left( 		m_\tau C^V_{RR} H_{V,+}     -  m_\tau  C^V_{LR}	H_{V,-}  		+ 4  		\sqrt{q^2}   	C^T_{RR}   H_{T2,-}   		\right) \nonumber \\
&+& \left(		2\sqrt{q^2} (C^S_{RR}-C^S_{LR}) H_S   + 		 2m_\tau 	(-C^V_{RR}+C^V_{LR}) H_{V,t}   		\right) \times \nonumber \\
&& \left(			(C^V_{LR}-C^V_{RR})^* \sqrt{q^2}  H_{V,0}    +  4  (C^T_{RR})^*  m_\tau  H_{T,0}  		\right) \nonumber \\
&+&  \left(			\sqrt{q^2} (C^V_{LR})^* H_{V,+}     -  \sqrt{q^2} (C^V_{RR})^*  H_{V,-}  	   +  4 m_\tau  			(C^T_{RR})^* H_{T2,+} 			\right)  \times \nonumber \\	
&& \left. \left(			m_\tau C^V_{LR}	 H_{V,+}    	- m_\tau C^V_{RR}		H_{V,-}     +  4   \sqrt{q^2}  C^T_{RR}		H_{T2,+} 			\right) \right] \nonumber 
\end{eqnarray}

Let us now move on to the forward-backward asymmetries $\mathcal{A}_{FB}^{(*)}$. Here we report the analytic results for a finer observable, namely the forward-backward asymmetry of $\tau$ with a specific helicity. For the $\bar{B}\rightarrow D \tau \nu$ decay involving the LH neutrinos we have 
\begin{eqnarray}
 {d {\mathcal A}_{FB}^{+} \over dq^2} &=& -\frac{G_F^2 V_{cb}^2}{192 m_B^3 \pi^3 q^2 \Gamma_{B\rightarrow D \tau \nu} }   \sqrt{\left(	(m_B-m_D)^2-q^2 \right) \left(	(m_B+m_D)^2-q^2 \right)  } \nonumber \\
\label{eq:analyticRDLAFBpl} 
 && (m_\tau^2 - q^2)^2 \frac{3}{2q^2}  \mathcal{R}e \left[ \left( \sqrt{q^2} (C^S_{RL}+C^S_{LL})	H_{S}^s 	+  	m_\tau 	(1+C^V_{LL}+C^V_{RL}) H_{V,t}^s	\right) \right. \nonumber \\
&& \left. \left(	 	m_\tau (1+C^V_{LL}+C^V_{RL})	H_{V,0}^s 		- 4 C^T_{LL} \sqrt{q^2} H^s_{T} 	\right)^*  \right] \nonumber  , \\
\\
{d {\mathcal A}_{FB}^{-}  \over dq^2}&=& 0. \nonumber 
\end{eqnarray}
Here the superscripts $\pm$ refer to specific $\tau$ helicities. Similarly, for the right-handed neutrinos contribution we have
\begin{eqnarray}
\label{eq:analyticRDRAFBpl} 
{d{\mathcal A}_{FB}^{+} \over dq^2} &=& 0 \nonumber \\ \\
 {d {\mathcal A}_{FB}^{- } \over dq^2} &=& -\frac{G_F^2 V_{cb}^2}{192 m_B^3 \pi^3 q^2 \Gamma_{B\rightarrow D \tau \nu}}   \sqrt{\left(	(m_B-m_D)^2-q^2 \right) \left(	(m_B+m_D)^2-q^2 \right)  } \nonumber \\
 && (m_\tau^2 - q^2)^2 \frac{3}{2q^2} \mathcal{R}e \left[ \left( \sqrt{q^2} (C^S_{RR}+C^S_{LR})	H_{S}^s + 	 	m_\tau  (C^V_{LR}+C^V_{RR}) H_{V,s}^s	\right) \right. \nonumber \\
&&  \left. \left(	 	m_\tau (C^V_{LR}+C^V_{RR})	H_{V,0}^s  		- 4 C^T_{RR} \sqrt{q^2} H^s_{T}   	\right)^* \right] \nonumber .
\end{eqnarray}

Equivalently, for the decays into $D^*$ we have
\begin{eqnarray}
\label{eq:analyticRDsLAFBplmin} 
{d {\mathcal A}_{FB}^{+*} \over dq^2}&=& -\frac{G_F^2 V_{cb}^2}{192 m_B^3 \pi^3 q^2 \Gamma_{B\rightarrow D^* \tau \nu} }   \sqrt{\left(	(m_B-m_{D^*})^2-q^2 \right) \left(	(m_B+m_{D^*})^2-q^2 \right)  } (m_\tau^2 - q^2)^2  \nonumber  \\
 &&  \frac{3}{2q^2}   \mathcal{R}e \left[ \left( \sqrt{q^2} (C^S_{RL}-C^S_{LL})	H_{S} 	+ 	 	m_\tau (1+C^V_{LL}-C^V_{RL}) H_{V,t}	\right)  \right. \nonumber \\
&& \left. \left( 	m_\tau  	(1+C^V_{LL}-C^V_{RL})	H_{V,0}		- 4 C^T_{LL} \sqrt{q^2} H_{T,0}	\right)^* \right]  , \nonumber\\ 
\\
{d {\mathcal A}_{FB}^{-*} \over dq^2} &=& \frac{G_F^2 V_{cb}^2}{192 m_B^3 \pi^3 q^2 \Gamma_{B\rightarrow D^* \tau \nu}  }   \sqrt{\left(	(m_B-m_{D^*})^2-q^2 \right) \left(	(m_B+m_{D^*})^2-q^2 \right)  }(m_\tau^2 - q^2)^2     \left(\frac{3}{4q^2} \right)    \nonumber \\
 &&  \mathcal{R}e \left[  \left( \sqrt{q^2} (1+C^V_{LL}-C^V_{RL})	(H_{V,-} + H_{V,+}) + 4 	m_\tau  C^T_{LL} (H_{T,-}	- H_{T,+}	)\right)   \right.   \nonumber \\
&&  \left. \left(  	 \sqrt{q^2} (1+C^V_{LL}+C^V_{RL})	(H_{V,-}- H_{V,+})	 + 4 	m_\tau  C^T_{LL} (H_{T,-}	 + H_{T,+}	) 	\right)^*  \right],  \nonumber 
\end{eqnarray}
for the LH neutrinos contribution. For the RH neutrino contribution we have
\begin{eqnarray}
\label{eq:analyticRDsRAFBpl} 
 {d {\mathcal A}_{FB}^{+* } \over dq^2} &=& -\frac{G_F^2 V_{cb}^2}{192 m_B^3 \pi^3 q^2 \Gamma_{B\rightarrow D^* \tau \nu} }   \sqrt{\left(	(m_B-m_{D^*})^2-q^2 \right) \left(	(m_B+m_{D^*})^2-q^2 \right)  } (m_\tau^2 - q^2)^2  \frac{3}{4q^2}  \nonumber \\
 &&    \mathcal{R}e \left[   \left( \sqrt{q^2} (C^V_{LR}-C^V_{RR})	(H_{V,-} + H_{V,+} ) 	+ 	4  	m_\tau  C^T_{RR} (- H_{T2,-} + H_{T2,+})\right)  \right. \nonumber \\
&& \left. \left(	 \sqrt{q^2} (C^V_{LR}+C^V_{RR})	(H_{V,-}  - H_{V,+} )	 - 	4  	m_\tau C^T_{RR} (H_{T2,-}	+ H_{T2,+}	) \right)^* \right] \nonumber , \\ 
\\
{d {\mathcal A}_{FB}^{-* } \over dq^2} &=& -\frac{G_F^2 V_{cb}^2}{192 m_B^3 \pi^3 q^2 \Gamma_{B\rightarrow D^* \tau \nu} }   \sqrt{\left(	(m_B-m_{D^*})^2-q^2 \right) \left(	(m_B+m_{D^*})^2-q^2 \right)  }  (m_\tau^2 - q^2)^2    \nonumber \\
 && \frac{3}{2q^2}  \mathcal{R}e \left[   \left( \sqrt{q^2} (C^S_{RR}-C^S_{LR})	H_{S} + 	m_\tau (C^V_{LR}-C^V_{RR}) H_{V,t} \right) \right. \nonumber \\
&& \left. \left(	 	m_\tau (C^V_{LR}-C^V_{RR})	H_{V,0} 	+ 4 C^T_{RR} \sqrt{q^2} H_{T,0} 	\right)^* \right]  \nonumber .
\end{eqnarray}
We note that $\mathcal{A}^{\pm,(*)}_{FB}$ contain more information than the $\mathcal{A}^{(*)}_{FB}$ observables we studied in this work. The experimental proposal in \cite{Alonso:2017ktd} for measuring the forward-backward asymmetry is applicable to the total asymmetry summed over final $\tau$ helicity. It is intriguing to find a similar proposal for measurement of $\mathcal{A}^{\pm,(*)}_{FB}$. In particular, \eqref{eq:analyticRDLAFBpl} and \eqref{eq:analyticRDRAFBpl} suggest that a non-zero $\mathcal{A}^{-}_{FB}$ ($\mathcal{A}^{+}_{FB}$) is a clear signature of RH (LH) neutrinos.

The numerical formulas in Sec.~\ref{sec:obsevables} are obtained by using similar form factors as in \cite{Bardhan:2016uhr}, and integrating the analytic formulas in this appendix over $q^2$. We use the same numerical parameter values as in \cite{Asadi:2018wea}, which we list here again in Tab.~\ref{tab:numbers} for completeness.

\begin{table*}[t]
\begin{tabular}{|c|c|c|c|}
\hline 
$V_{cb}$ & $G_F$ [GeV$^{-2}$] & $m_{\bar{B}}$ [GeV] & $m_D$ [GeV] \\ 
\hline 
$42.2 \times 10^{-3}$ & $1.166 \times 10^{-5}$ & 5.279 & 1.870  \\ 
\hline 
\hline 
$m_{D^*}$ [GeV] & $m_e$ [GeV] & $m_\mu$ [GeV] & $m_\tau$ [GeV] \\
\hline 
2.010 & $0.511 \times 10^{-3}$ & 0.106 & 1.777 \\
\hline 
\end{tabular} 
\caption{The parameter values used in the calculation of the  numerical formulas in this paper.} 
\label{tab:numbers}
\end{table*}

\section{Scanning the Parameter Space of LQ Models}
\label{sec:scan}

In this appendix we describe a numerical method to calculate all the CP-even asymmetry observables as a function of $R_{D^{(*)}}$ and $Br \left( B_c \rightarrow \tau \nu \right)$. This will allow us to efficiently carry out a scan over the parameter space of different models that generate the correct value of $R_{D^{(*)}}$ and respect the bound \eqref{eq:Bcbound} on $Br \left( B_c \rightarrow \tau \nu \right)$. 

For models that result in a single Wilson coefficient, we simply scan over the two degrees of freedom (the real and imaginary coefficient values) and find the ones respecting the conditions on $R_{D^{(*)}}$ and $B_c$ decay rate. 

For the LQ mediators that result in two independent complex coefficients ($S_1$ and $U_1$ LQs), we have a scalar current Wilson coefficient that we denote by $C^S$, and a vector current denoted by $C^V$. When these LQs are coupled to LH neutrinos, namely the $S_1^L$ and the $U_1^L$ models, we absorb the SM contribution into $C^V$ as well. Given the CP-even property of $R_{D^{(*)}}$ observables and $Br \left( B_c \rightarrow \tau \nu \right)$, we can write
\begin{eqnarray}
\label{eq:appx3WC}
R_{D^{(*)}} & = & x^m_{R_{D^{(*)}}}|C^S|^2 + y^m_{R_{D^{(*)}}} |C^V|^2 + z^m_{R_{D^{(*)}}} \mathcal{R}e \left[	C^S C^V \right]+ w^m_{R_{D^{(*)}}}, \\
Br \left( B_c \rightarrow \tau \nu \right) & = & x^m_{B_c}|C^S|^2 + y^m_{B_c} |C^V|^2 + z^m_{B_c} \mathcal{R}e \left[	C^S C^V \right]+ w^m_{B_c}, \nonumber 
\end{eqnarray}
where $m=S_1^L, U_1^L, U_1^R$ stands for the model under study. The coefficients $x^m_{R_{D^{(*)}}}$, $y^m_{R_{D^{(*)}}}$, $z^m_{R_{D^{(*)}}}$, and $w^m_{R_{D^{(*)}}}$ are real numbers. Similarly, for the numerator of all the other CP-even observables we have
\begin{equation}
R_{D^{(*)}} O^{(*)}  =  x^m_{O^{(*)}}|C^S|^2 + y^m_{O^{(*)}} |C^V|^2 + z^m_{O^{(*)}} \mathcal{R}e \left[	C^S C^V \right]+ w^m_{O^{(*)}}, 
\label{eq:appx3WCobs}
\end{equation}
where $O^{(*)} $ stands for $\mathcal{A}_{FB}^{(*)}$, $\mathcal{P}_{\tau}^{(*)}$, or $\mathcal{P}_{\perp}^{(*)}$.

For each model, the numerical coefficients $\left( x^m_O, y^m_O, z^m_O, w^m_O \right)$ can be calculated from the numerical formulas \eqref{eq:numericRDAFB} and \eqref{eq:numericPtau}--\eqref{eq:numericPperp} for each observable. We see that although we have two independent complex Wilson coefficients, only 3 real degrees of freedom (here $|C^V|$, $|C^S|$, and $\mathcal{R}e \left[	C^S C^V \right]$) appear (linearly) in the equations. We can invert these linear equations and express all the observables in terms of any three of them. For the purposes of performing a scan, we choose to express all the observables in terms of $R_D$, $R_{D^*}$, and $Br \left( B_c \rightarrow \tau \nu \right)$:
\begin{equation}
R_{D^{(*)}} O^{(*)} = a^{O^{(*)}}_{m} R_D + b^{O^{(*)}}_{m} R_{D^*} + c^{O^{(*)}}_{m} Br \left( B_c \rightarrow \tau \nu \right) + d^{O^{(*)}}_{m},
\label{eq:linearappx}
\end{equation}
where the coefficients $a^{O^{(*)}}_{m}$, $b^{O^{(*)}}_{m}$, $c^{O^{(*)}}_{m}$, $d^{O^{(*)}}_{m}$ are real numbers. We can calculate these coefficients once and for all for each model and observable, and then use \eqref{eq:linearappx} to find each observable from $R_{D^{(*)}}$ and $Br \left(	B_c \rightarrow \tau \nu	\right)$. This allows us to perform a highly efficient scan over the entire viable parameter space and 
 calculate all the CP-even observables.

\afterpage{\clearpage}

\bibliographystyle{utphys_modified}
\bibliography{bib}

\end{document}